\documentclass{aastex63}
\usepackage{float}
\received{\today}
\submitjournal{ApJ}

\shorttitle{Photochemistry of Anoxic Abiotic Habitable Planets}
\shortauthors{Ranjan et al.}

\begin{document}

\title{Photochemistry of Anoxic Abiotic Habitable Planet Atmospheres: Impact of New H$_2$O Cross-Sections}

\correspondingauthor{Sukrit Ranjan}
\email{sukrit@mit.edu}

\author[0000-0002-5147-9053]{Sukrit Ranjan}
\affiliation{Massachusetts Institute of Technology \\
77 Massachusetts Avenue\\
Cambridge, MA 02140, USA}
\affiliation{Northwestern University \\
1800 Sherman Ave\\
Evanston, IL 60201}
\affiliation{SCOL Postdoctoral Fellow}

\author[0000-0002-2949-2163]{Edward W. Schwieterman}
\affiliation{School of Earth and Atmospheric Sciences\\
Georgia Institute of Technology}
\affiliation{Department of Earth and Planetary Sciences\\
University of California, Riverside}
\affiliation{Blue Marble Space Institute of Science\\
Seattle, WA}

\author[0000-0003-2281-1990]{Chester Harman}
\affiliation{NASA Ames Research Center}

\author[0000-0003-2863-2707]{Alexander Fateev}
\affiliation{Technical University of Denmark\\
Department of Chemical and Biochemical Engineering\\
Frederiksborgvej 399, Roskilde\\
DK 4000, Denmark}

\author[0000-0002-7853-6871]{Clara Sousa-Silva}
\affiliation{Massachusetts Institute of Technology \\
77 Massachusetts Avenue\\
Cambridge, MA 02140, USA}

\author{Sara Seager}
\affiliation{Massachusetts Institute of Technology \\
77 Massachusetts Avenue\\
Cambridge, MA 02140, USA}

\author[0000-0003-2215-8485]{Renyu Hu}
\affiliation{Jet Propulsion Laboratory\\ California Institute of Technology\\ Pasadena, CA 91109, USA}
\affiliation{Division of Geological and Planetary Sciences\\ California Institute of Technology\\ Pasadena, CA 91125, USA}



\begin{abstract}


We present a study of the photochemistry of abiotic habitable planets with anoxic CO$_2$-N$_2$ atmospheres. Such worlds are representative of early Earth, Mars and Venus, and analogous exoplanets. H$_2$O photodissociation controls the atmospheric photochemistry of these worlds through production of reactive OH, which dominates the removal of atmospheric trace gases. The near-UV (NUV; $>200$ nm) absorption cross-sections of H$_2$O play an outsized role in OH production; these cross-sections were heretofore unmeasured at habitable temperatures ($<373$ K). We present the first measurements of NUV H$_2$O absorption at $292$ K, and show it to absorb orders of magnitude more than previously assumed. To explore the implications of these new cross-sections, we employ a photochemical model; we first intercompare it with two others and resolve past literature disagreement. The enhanced OH production due to these higher cross-sections leads to efficient recombination of CO and O$_2$, suppressing both by orders of magnitude relative to past predictions and eliminating the low-outgassing ``false positive" scenario for O$_2$ as a biosignature around solar-type stars. Enhanced [OH] increases rainout of reductants to the surface, relevant to prebiotic chemistry, and may also suppress CH$_4$ and H$_2$; the latter depends on whether burial of reductants is inhibited on the underlying planet, as is argued for abiotic worlds. While we focus on CO$_2$-rich worlds, our results are relevant to anoxic planets in general. Overall, our work advances the state-of-the-art of photochemical models by providing crucial new H$_2$O cross-sections and resolving past disagreement in the literature, and suggests that detection of spectrally active trace gases like CO in rocky exoplanet atmospheres may be more challenging than previously considered. 
\end{abstract}



\section{Introduction}
The statistical finding that rocky, temperate exoplanets are common \citep{Dressing2015} has received dramatic validation with the discovery of nearby potentially-habitable worlds like LHS-1140b, TRAPPIST-1e, TOI-700d, and Kepler-442b \citep{Dittmann2017, Gillon2017, Gilbert2020, Torres2015}. Upcoming facilities such as the James Webb Space Telescope (JWST), the Extremely Large Telescopes (ELTs), and the HabEx and LUVOIR mission concepts will have the ability to detect the atmospheres of such worlds, and possibly characterize their atmospheric compositions \citep{Rodler2014, Fujii2018, Lustig-Yaeger2019, LUVOIR2019, Meixner2019, Gaudi2020}. 

The prospects for rocky exoplanet atmospheric characterization have lead to extensive photochemical modelling of their potential atmospheric compositions, with emphasize on constraining the possible concentrations of spectroscopically active trace gases like CO, O$_2$, and CH$_4$. Particular emphasis has been placed on modelling the atmospheres of habitable but abiotic planets with anoxic, CO$_2$-N$_2$ atmospheres (e.g., \citealt{Segura2007, Hu2012, Rugheimer2015, Rimmer2016, Schwieterman2016, Harman2018, James2018, Hu2020}). Such atmospheres are expected from outgassing on habitable terrestrial worlds, are representative of early Earth, Mars and Venus \citep{Kasting1993, Wordsworth2016, Way2016}, and are expected on habitable exoplanets orbiting younger, fainter stars or at the outer edges of their habitable zones (e.g., TRAPPIST-1e; \citealt{Kopparapu2013hz, Wolf2017}). 

Water vapor plays a critical role in the photochemistry of such atmospheres, because in anoxic abiotic atmospheres, H$_2$O photolysis is the main source of the radical OH, which is the dominant sink of most trace atmospheric gases \citep{Rugheimer2015mdwarf, Harman2015}. Most water vapor is confined to the lower atmosphere due to the decline of temperature with altitude and the subsequent condensation of H$_2$O, the ``cold trap" (e.g., \citealt{Wordsworth2013CO2}). In CO$_2$-rich atmospheres, this abundant lower atmospheric H$_2$O is shielded from UV photolysis at FUV wavelengths ($\leq200$ nm)\footnote{The partitioning of the UV into near-UV (NUV), far-UV (FUV), and sometimes mid-UV (MUV) is highly variable (e.g., \citealt{France2013, Shkolnik2014, Domagal-Goldman2014, Harman2015}). In this paper, we adopt the nomenclature of \citet{Harman2015} that NUV corresponds to $>200$ nm and FUV to $\leq200$ nm, because this partitioning approximately coincides with the onset of CO$_2$ absorption.}, meaning that NUV ($>200$ nm) absorption plays an overweight role in H$_2$O photolysis \citep{Harman2015}. Therefore, the NUV absorption cross-sections of H$_2$O are critical inputs to photochemical modelling of anoxic abiotic habitable planet atmospheres. However, to our knowledge, the NUV absorption of H$_2$O at habitable temperatures ($<373$ K) was not known  prior to this work. In the absence of measurements, photochemical models relied upon varying assumptions regarding H$_2$O NUV absorption \citep{Kasting1981, Sander2011, Rimmer2016, Rimmer2019}. 

In this paper, we present the first-ever measurements of the NUV cross-sections of H$_2$O(g) at temperatures relevant to habitable worlds ($T=292\text{K}<373$ K), and explore the implications for the atmospheric photochemistry of abiotic habitable planets with anoxic CO$_2$-N$_2$ atmospheres orbiting Sun-like stars. We begin by briefly reviewing the photochemistry of anoxic CO$_2$-N$_2$ atmospheres, and the key role of H$_2$O (Section~\ref{sec:CO_background}). We proceed to measure the cross-sections of H$_2$O(g) in the laboratory, and find it to absorb orders of magnitude more in the NUV than previously assumed by any model; this laboratory finding is consistent with our (limited) theoretical understanding of the behaviour of the water molecule (Section~\ref{sec:measurements}). We incorporate these new cross-sections into our photochemical model. Previous models of such atmospheres have been discordant; we conduct an intercomparison between three photochemical models to successfully reconcile this discordance to within a factor of $2\times$ (Section~\ref{sec:model_intercompare}). We explore the impact of our newly measured, larger H$_2$O cross-sections and their concomitantly higher OH production on the atmospheric photochemistry and composition for our planetary scenario (Section~\ref{sec:updated_model}). We focus on O$_2$ and especially CO, motivated by their spectral detectability and proposed potential to discriminate the presence of life on exoplanets \citep{Snellen2010, Brogi2014, Rodler2014, Wang2016, Schwieterman2016, Meadows2018, Krissansen-Totton2018, Schwieterman2019}, but we consider the implications for other species as well, and especially CH$_4$ and H$_2$. We summarize our findings in Section~\ref{sec:disc_conc}. The Appendices contain supporting details: Appendix~\ref{sec:detailed_simulation_parameters} details our simulation parameters, Appendix~\ref{sec:detailed_boundary_conditions} details the boundary conditions, and Appendix~\ref{sec:detailed_model_intercomparison} details our model intercomparison and the insights derived thereby. While we focus here on CO$_2$-rich atmospheres, our results are relevant to any planetary scenario in which H$_2$O photolysis is the main source of OH, which includes most anoxic atmospheric scenarios.


\section{Photochemistry of CO$_2$-Rich Atmospheres \label{sec:CO_background}}
In this section, we briefly review the photochemistry of CO$_2$-rich atmospheres. For a more detailed discussion, we refer the reader to \citet{CatlingKasting2017}.

UV light readily dissociates atmospheric CO$_2$ via $CO_2+ h\nu \rightarrow CO + O$ ($\lambda<202$ nm; \citealt{ityaksov2008co2}). The direct recombination of CO and O is spin forbidden and slow, and the reaction $O + O + M \rightarrow O_2 + M$ is faster. Consequently, CO$_2$ on its own is unstable to conversion to CO and O$_2$ \citep{Schaefer2011}. However, OH can react efficiently with CO via reaction $CO + OH \rightarrow CO_2+H$, and is the main photochemical control on CO and O$_2$ via catalytic cycles such as:
\begin{eqnarray}
    CO + OH \rightarrow CO_2 + H\\
    O_2 + H + M \rightarrow HO_2 + M\\
    HO_2+O \rightarrow O_2 + OH\\
    ---------------------------\nonumber \\
    Net: CO + O \rightarrow CO_2 
\end{eqnarray}

On Mars, such OH-driven catalytic cycles stabilize the CO$_2$ atmosphere against conversion to CO and O$_2$ \citep{McElroy1972, Parkinson1972, Krasnopolsky2011}. These catalytic cycles are diverse, but unified in requiring OH to proceed \citep{Harman2018}. Similarly on modern Earth, OH is the main sink on CO \citep{Badr1995}. On Venus, the products of HCl photolysis are thought to support this recombination \citep{Prinn1971, Yung1982, Mills2007, Sandor2018}; however, this mechanism should not be relevant to habitable planets with hydrology, where highly soluble HCl should be efficiently scrubbed from the atmosphere \citep{Lightowlers1988, Prinn1987}.

On oxic modern Earth, the main source of OH is the reaction $O(^{1}D)+H_2O\rightarrow 2OH$, with $O(^{1}D)$ sourced from O$_3$ photolysis via $O_3+h\nu \rightarrow O(^{1}D)+O_2$ ($\lambda<320$ nm; \citealt{Jacob1999}). However, on anoxic worlds, O$_3$ is low, and OH is instead ultimately sourced from H$_2$O photolysis ($H_2O+h\nu\rightarrow OH+H$), though it may accumulate in alternative reservoirs \citep{Tian2014, Harman2015}. Consequently, on anoxic abiotic worlds, the balance between CO$_2$ and H$_2$O photolysis is thought to control the photochemical accumulation of CO in the atmosphere. OH also reacts with a wide range of other gases. Hence, the photochemistry of CO$_2$-dominated atmospheres is controlled by H$_2$O, through its photolytic product OH. 

The proper operation of this so-called HO$_X$ photochemistry in photochemical models is commonly tested by reproducing the atmosphere of modern Mars, which is controlled by these processes (e.g., \citealt{Hu2012}). However, the atmosphere on modern Mars is thin ($\sim0.006$ bar), whereas the atmospheres of potentially habitable worlds are typically taken to be more Earth-like ($\sim1$ bar). As we will show, models which are convergent in the thin atmospheric regime of modern Mars may become divergent for thicker envelopes, illustrating the need for intercomparisons in diverse regimes to assure model accuracy (Section~\ref{sec:model_intercompare}). In particular, on planets with high CO$_2$ abundance, the H$_2$O-rich lower atmosphere is shielded from FUV radiation by CO$_2$, meaning that H$_2$O photolysis at low altitudes is dependent on NUV photons.

In addition to the atmospheric sources and sinks discussed here, CO may have strong surface sources and sinks. In particular, impacts, outgassing from reduced melts, and biology may supply significant CO to the atmosphere, and biological uptake in the oceans may limit [CO] in some scenarios \citep{Kasting1990, Kharecha2005, Batalha2015, Schwieterman2019}. In this work, we focus solely on photochemical CO, and neglect these other sources and sinks.

\section{Measurements of NUV H$_2$O Cross-Sections at 292K\label{sec:measurements}}

As discussed above, NUV H$_2$O photolysis is critical to the photochemistry of abiotic habitable planets with anoxic CO$_2$-rich atmospheres. However, prior to this work, no experimentally measured or theoretically predicted absorption cross-sections were available for H$_2$O(g) at habitable conditions ($T<373$ K) at wavelengths $>198$ nm \citep{Burkholder2015}. This is because H$_2$O absorption cross-sections are very low ($<1\times10^{-20}$ cm$^{2}$ at $\geq190$ nm) that make their measurement difficult. Here, we extend this coverage to 230 nm, by measuring the absorption cross-section of H$_2$O(g) at 292 K between 186 and 230 nm (0.11 nm spectral resolution). We describe our method (Section~\ref{sec:new_h2o_exp}), consider the consistency with past measurements and theoretical expectations, and prescribe H$_2$O cross-sections for inclusion into atmospheric models (Section~\ref{sec:new_h2o_theory_comp}). 

\subsection{Experimental Set-Up And Measurements\label{sec:new_h2o_exp}}
The measurements have been performed in a special flow gas cell. The cell is made from a stainless steel tube ($\sim$25mm inner diameter) in a straight-line design and is 570 cm long. The cell is thermally isolated and it can be heated up to $200^\circ$C. Exchangeable sealed optical widows at the both ends allow optical measurements in a wide spectral range from far-UV to far-IR (defined by window material). In the present measurements MgF$_2$ VUV windows have been used.

The cell is coated inside with SilcoNert 2000 coating\footnote{Silcotek Company: \url{ https://www.silcotek.com/silcod-technologies/silconert-inert-coating}} which has good hydrophobic properties\footnote{Silcotek hydrophobicity rating of ``3"} and is very inert\footnote{Silcotek chemical inertness rating of ``4"} to various reactive gases allowing low-level optical absorption measurements (e.g. sulfur/H$_2$S, NH$_3$, formaldehyde etc.) in various laboratory and industrial environments (e.g. analytical, stack and process gases). 

Flow through the cell is controlled with a high-end mass-flow controller (MFC) (BRONKHORST). The pressure measurements in the cell were calibrated with a high-end ROSEMOUNT pressure sensor. 

Near-UV absorption measurements were done with use a 0.5 m far-UV spectrometer equipped with X-UV CCD (Princeton Instruments) (spectral range $110-240$ nm), far-UV coated collimating optics (mirrors) and VUV D2-lamp (HAMAMATSU). The spectrometer and the optics were purged with N$_2$ (99.999\%). Because H$_2$O has continuum-like absorption in $180-240$ nm the measurements have been done with 600 grooves mm$^{-1}$ grating blazing at 150 nm without spectral scanning (spectral resolution $\Delta \lambda=0.11$ nm).

For H$_2$O measurements a gas-tight HAMILTON syringe\footnote{Hamilton Company: \url{https://www.hamiltoncompany.com/laboratory-products/syringes/general-syringes/gastight-syringes/1000-series}} and an accurate syringe pump with a water evaporator (heated to $150^\circ$C) were used in order to produce controlled N$_2$+H$_2$O (1.5–2\%) mixtures. Milli-Q water\footnote{\url{https://dk.vwr.com/store/product/en/2983107/ultrarene-vandsystemer-milli-q-reference?languageChanged=en}}, purified from tap water, was used. We did not characterize the isotopic composition of our tap water, but see no reason for the heavier isotopes of water to be absent. This means that our water vapor cross-sections should include contributions from heavier isotopes of water, such as HDO and D$_2$O. Since these heavier isotopes should be present on other planets as well, we argue our use of tap water to be appropriate for representative spectra of water vapor for planetary simulations. As a practical note, the finding of \citet{Chung2001} that $\sigma_{H_{2}O}$ is significantly larger than $\sigma_{HDO}$ and $\sigma_{D_{2}O}$ for $\lambda>180$ nm, combined with the low absolute abundances of these heavier isotopes, indicates their contribution to our measured NUV spectrum should be minimal. The water evaporation system (syringe + pump + evaporator) was the same as previously used in high-temperature N$_2$+H$_2$O transmissivity measurements \citep{Ren2015}. 

In the absorption measurements cold N$_2$ (i.e. at 19$^\circ$C) flows into a heated evaporator where H$_2$O from a syringe pump is mixed in. The N$_2$ + H$_2$O mixture then enters a 2 meter unheated Teflon line (inner diameter 4mm), connecting to the cell, where the mixture naturally cools down. N$_2$ flow through the system was kept at 2 $l_n$ min$^{-1}$ in all measurements ($l_n$ min$^{-1}$ = normal liter per minute). Effective residence time of the gas in the cell at that flow rate and 19.4$^\circ$C was 1.31 min.

To account for any heat-transfer effects from the injection of cold N$_2$ through the heated evaporator and into the cell, reference measurements (i.e. without H$_2$O(g)) have been performed with N$_2$ at 99.999\%.  The outlet of the cell was kept open. Temperature in the cell was continuously measured in two zones with thermocouples. Temperature in the cell during the measurements was between 19.2$^\circ$C and 19.7$^\circ$C with temperature uniformity $\pm~0.1^\circ$C at a particular measurement. It should be noted that H$_2$O saturation point at 19.2$^\circ$C is 2.19 volume \% at the conditions of the measurements. Therefore all measurements with water were below saturation conditions. Prior to our measurements, we purged our apparatus with dry air (H$_2$O and CO$_2$-free) for $\sim 2$ days.

We conducted four measurement sequences of the H$_2$O cross-sections. In each of the first three measurement sequences, we began by taking a reference spectrum of dry N$_2$. We then injected 1.5\%water vapor into our apparatus, and took 5-6 spectra of N$_2$+1.5\% H$_2$O. We finished the sequence with another N$_2$ spectrum for a baseline check. In the fourth measurement sequence, we instead injected 2\% water vapor into our apparatus, in order to get a better signal-to-noise (S/N) ratio in the 215-230 nm wavelength where H$_2$O has the lowest absorption cross-sections. The first of the 5-6 N$_2$+H$_2$O measurements in a sequence was performed to ensure that [H$_2$O] was stable; it was discarded and did not contribute towards the cross-section calculations. We also discarded the first of the four sequences because the measurements were not stable; we attribute this to H$_2$O saturating the surfaces of the dry system, which had been purged for days. Consequently, the final mean spectrum is based on 3 sequences of 4-5 measurements each, 2 at 1.5\% H$_2$O and 1 at 2.0\% H$_2$O. 

Absorption cross-sections were calculated assuming Lambert-Beer law: 

\begin{eqnarray} 
\label{eqn:beerlambert}
    \tau(\lambda)=\ln\frac{I_{0}(\lambda)}{I_{1}(\lambda)} \\
    \sigma(\lambda)=\frac{\tau(\lambda)}{n_{H_{2}O}L}
\end{eqnarray}

where
\begin{itemize}
    \item $\lambda$ is wavelength, specified in nm in our apparatus;
    \item $L$ is the path length in cm (L=570 cm in our apparatus);
    \item $n_{H_{2}O}$ is water concentration in molecules cm$^{-3}$(defined by amount of evaporated water, temperature and pressure in the cell);
    \item $\sigma(\lambda)$ is the absorption cross-section in cm$^2$ molecule$^{-1}$;
    \item $\tau(\lambda)$ is the optical depth;
    \item $I_{0}(\lambda)$ is the reference spectrum (99.999\% N$_2$ in the cell);
    \item $I_{1}(\lambda)$ is the absorbance spectrum (99.999\% N$_2$ $+$ 1.5\% or 2\% H$_2$O in the cell).
\end{itemize}

Both $I_{0}(\lambda)$ and $I_{1}(\lambda)$ are corrected for stray light in the system. 


We averaged the measured cross-sections from our 3 sequences of 4-5 measurements each to calculate a mean absorption spectrum and estimate its errors. Absorption cross-sections are calculated with use of Equation~\ref{eqn:beerlambert}. Errors of the mean absorption cross-sections are calculated taking into account the mean experimental standard deviations of the absorption spectra and standard deviations in pressure and water concentrations measurements and calculations. The former is defined by the pressure sensor used and the latter is defined by uncertainties in N$_2$ flow (MFC) and water evaporation system. Combined uncertainties in pressure and water concentrations were $\pm0.77\%$ (for 1.5\% H$_2$O) and $\pm0.74\%$ (for 2\% H$_2$O). This means that absolute uncertainty in water concentrations in the N$_2$ carry gas are ($1.50\pm0.01$) and ($2.00\pm0.02$) volume \%. Temperature variations in the cell are negligible in the uncertainty calculations. 

\begin{figure}[h]
\plotone{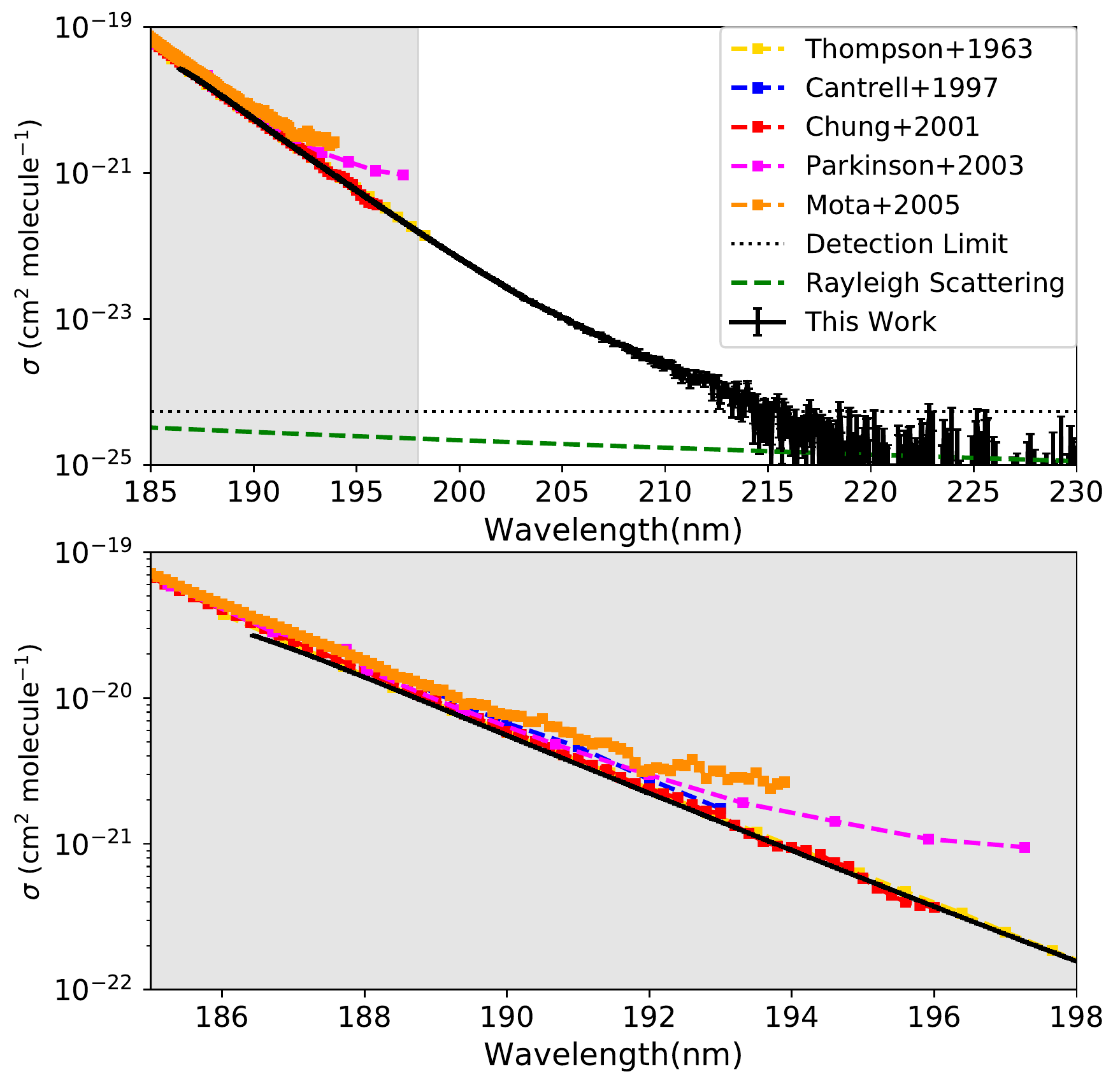}
\caption{New H$_2$O absorption cross-sections compared to published data. Also plotted is the Rayleigh scattering cross-section, calculated as in \citet{Ranjan2017a}. The top plot shows the full dataset; the bottom zooms in on the 185-198 nm wavelength region (highlighted in grey on the top plot) where data for comparison is available. The dotted black line demarcates the detection limit of our apparatus (Section~\ref{sec:new_h2o_exp}). \label{fig:new_h2o_xc}}
\end{figure}

The composite absorption cross-sections in 186.45-230.413 nm are calculated by calculating the mean cross-sections data from the last three sequences and their respective standard deviations, which are taken to represent the $1-\sigma$ error assuming Gaussian statistics. The resulting cross-sections are shown in Figure~\ref{fig:new_h2o_xc}. The minimum measurable absorption in our experimental apparatus, calculated by the ratio of incident to detected light is $1.53\times{10^{-4}}$, corresponding to a minimum measurable cross section of $5.41\times{10^{-25}}$ cm$^2$ molecule$^{-1}$. 

\subsection{Comparison to Previous Data \& Theoretical Expectations \label{sec:new_h2o_theory_comp}}

In this section, we compare our measurements (Figure~\ref{fig:new_h2o_xc}) to previously published data and to theoretical expectations, to assess their quality and ascertain our confidence in them.

Data to compare our measurements to are nonexistent for $\lambda>198$ nm\footnote{A measurement at 207 nm was reported by \citet{Tan1978}, but the measurements from this dataset at $>180$ nm are in tension by orders of magnitude with all other measurements and are considered to be erroneous \citep{Chan1993h2o}; we therefore exclude it from consideration.}, but some datasets are available for $\lambda>190$ nm \citep{Thompson1963, Cantrell1997, Chung2001, Parkinson2003h2o, Mota2005}. Our data agree with the measurements of \citet{Thompson1963}, \citet{Cantrell1997}, and \citet{Chung2001}, with best agreement with the dataset of \citet{Chung2001}, which is the most conservative of all datasets (i.e. presumes the lowest water absorption for wavelengths between 190 and 198 nm).

Our data agree with \citet{Mota2005} and \citet{Parkinson2003} at shorter wavelengths, but disagree with these datasets at their red edges. At these red edges, both \citet{Mota2005} and \citet{Parkinson2003} show distinctive upturns in the water absorption at the long-wavelength edge of their measurements, in disagreement with both the expected behavior of the spectra, and \citet{Chung2001} and \citet{Thompson1963}. There are many possible explanations for such disagreement between sets of data, including experimental limitations (i.e. most of the disagreements occur at the instrumental threshold of measurements), variation in baseline corrections, and other experimental set-up concerns (e.g., whether equilibrium conditions in the system have been established before measurements, potential for underestimating water concentration, and scattering from H$_2$O in their saturated measurements). Of these, experimental limitations are a particularly compelling explanation, since the disagreements with \citet{Mota2005} and \citet{Parkinson2003} occur where one would expect them to be a problem, i.e. where their H$_2$O cross-sections are weakest and their measurement setups are closest to their limits. We therefore attribute the upturn at the red edges of the datasets of \citet{Mota2005} and \citet{Parkinson2003} to experimental error; we below apply this same logic to our own dataset.

Theoretical predictions expect that, at room temperature, water absorbs very weakly at wavelengths $\geq$180nm, losing intensity with a roughly exponential trend until it reaches the H-OH bond dissociation energy near 240 nm. This can be considered a vapor equivalent of Urbach's rule which predicts that, as an electronically excited band moves away from its peak, the absorption coefficient  decreases approximately as an exponential of the transition frequency \citep{Quickenden1980}.

As wavelengths increase towards dissociation, the populations of the energy levels participating in the transitions that cause the spectral absorption become increasingly sparse. This thermal occupancy factor is the strongest effect in predicting absorption in this region, and corresponds to an exponential decay of transition strength, which gives the cross-section its recognizable log-linear shape between 180 and 240 nm. The opacity in this region is caused by transitions to excited electronic states of water, which are effectively unbound even at their lowest energy. Consequently, other weak effects can provide minor contributions towards the total absorption that can lead to a small upturn in the overall spectrum. For example, pre-dissociation effects can broaden the wings of the hot, combination rovibronic bands in the region, resulting in a small gain in opacity. Additionally, Frank-Condon factors and Einstein-A coefficients, which are hard to predict in this region, can increase near dissociation and consequently limit the loss of line strength caused by the reduction in thermal occupancy of the transition states.

The new measured data presented here agrees with the qualitative theoretical expectations of the spectral behaviour of water in the wavelength range 186 - 215 nm. From 215 nm to 230 nm, our measured data exhibit an upward deviation from log-linear decrease, similar to the deviation reported in the long-wavelength edge of other previously measured data (e.g., \citealt{Parkinson2003, Mota2005}). This upward deviation in all three datasets corresponds to the region of the measurements that approaches the instrumental noise floor, which not only introduces uncertainty to each measured data point, but also to the overall predicted absorption. It is therefore not fully certain that this upward deviation is physical.

Given the sensitivity of photochemical models to increased water absorption in the 186-230 nm region, we have adapted the measured cross-sections presented here to minimize the possibility that our predicted water absorption is overestimated. To this end, we considered two prescriptions for $292$K H$_2$O(g) absorption cross-sections for incorporation into our photochemical models. The first prescription corresponds to our measured data with a cut-off after 216.328 nm, which is where the ratio of the measured absorption to the errors first goes below 3. We term this the ``cutoff" prescription. We note this is a conservative but unphysical prescription, since the water absorption at wavelengths above 216 nm is not expected to collapse, but instead experience an exponential loss in intensity. Our second data set addresses the concerns above by replacing the measured data at wavelengths $>$205 nm with a theoretical extrapolation, corresponding to a log-linear loss of absorption of our data from 186-205 nm towards dissociation (longer wavelengths). We term this the ``extrapolation" prescription. This prescription is similar in spirit to that executed by \citet{Kasting1981}, but with the advantage of the greater spectral coverage and higher sensitivity of our new dataset, which significantly affect the results. We note that this extrapolation is expected to underestimate overall opacity (see above for potential quantum chemical effects that can increase opacity near dissociation).

\begin{figure}[h]
\plotone{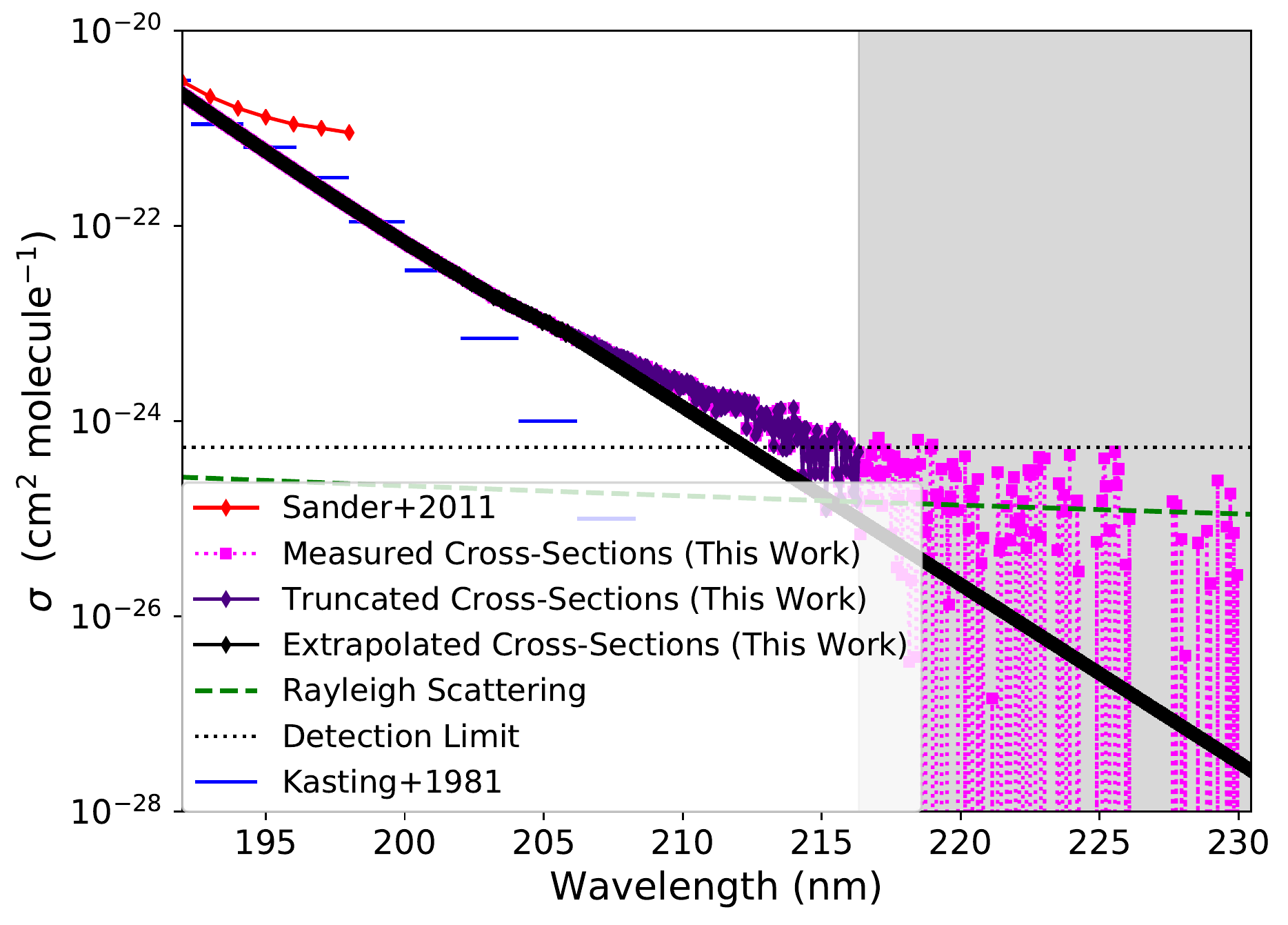}
\caption{Cutoff and extrapolation prescriptions for the spectral absorption cross-sections of 292K H$_2$O(g). Also shown are the recommended cross-sections of \citet{Sander2011}, the prescription of \citet{Kasting1981}, the Rayleigh scattering cross-sections \citep{Ranjan2017a}, and the full dataset of measured cross-sections reported here. Our measured cross-sections first dip below a $3-\sigma$ significance at 216.348 nm; in the spirit of conservatism we consider our data at $>216.348$ nm unreliable, and demarcate it as such with grey shading. The dotted black line demarcates the detection limit of our apparatus (Section~\ref{sec:new_h2o_exp}). \label{fig:new_h2o_xc_inmodels}}
\end{figure}

 Figure~\ref{fig:new_h2o_xc_inmodels} presents both of our prescriptions for 292K NUV H$_2$O(g) absorption cross-sections; also shown are the prescription of \citet{Kasting1981} and the recommended cross-sections of \citet{Sander2011}. At wavelengths $<192$ nm, we employ the recommended cross-sections of \citet{Sander2011} (i.e. we replace the \citealt{Parkinson2003} cross-sections of \citealt{Sander2011}). Both prescriptions specified here should be considered conservative choices, in that if anything they underestimate H$_2$O(g) absorption in this wavelength range. Nevertheless, both prescriptions indicate H$_2$O(g) absorption $\geq205$ nm to be orders of magnitude higher than previously considered, with profound photochemical implications (Section~\ref{sec:updated_model}). Both prescriptions, together with the underlying measurements, are available at \url{https://github.com/sukritranjan/ranjanschwietermanharman2020}.

\section{Model Intercomparison \label{sec:model_intercompare}}
We seek to determine the photochemical effects of our new H$_2$O cross-sections on the atmospheric composition of abiotic habitable worlds with anoxic CO$_2$-N$_2$ atmospheres. However, modelling of these atmospheres is discordant, with disagreement on a broad range of topics. Broadly, the models feature order-of-magnitude disagreements as to the trace gas composition of such atmospheres, and in particular their potential to accumulate photochemical CO and O$_2$ \citep{Kasting1990, Zahnle2008, Hu2012, Tian2014, Domagal-Goldman2014, Harman2015, Rimmer2016, James2018, Hu2020}.

To resolve this disagreement and derive a robust model for use in this work, we intercompare the models of \citet{Hu2012}, \citet{Harman2015}, and ATMOS (\citealt{Arney2016}, commit \#be0de64; \texttt{Archaean+haze} template). For convenience, we allude to the model of \citet{Hu2012} as the ``Hu model" and the model of \citet{Harman2015} as the ``Kasting model" to reflect their primary developers, with the caveat that multiple workers have contributed to these models. We apply these models to the CO$_2$-dominated benchmark planetary scenario outlined in \citet{Hu2012}. This scenario corresponds to an abiotic rocky planet orbiting a Sun-like star, with a 1-bar 90\% CO$_2$, 10\% N$_2$ atmosphere with surface temperature 288 K. Appendices~\ref{sec:detailed_simulation_parameters} and ~\ref{sec:detailed_boundary_conditions} present the details of the planetary scenario and boundary conditions adopted by these models. We focus on the surface mixing ratio of CO, $r_{CO}$, as the figure of merit for the intercomparison. At the outset of the intercomparison, the predictions of [CO] for this planetary scenario varied by $50\times$ between these models (Table~\ref{tab:intercomparison_results}). 

We describe in detail the key differences between our models which drove the disagreement in $r_{CO}$ in Appendix~\ref{sec:detailed_model_intercomparison}. Briefly, we identified the following errors and necessary corrections in our models (c.f. Appendix~\ref{sec:detailed_model_intercomparison}, Table~\ref{tab:corrections}):

\begin{itemize}
    \item \textbf{Correction of CO$_2$ Absorption Cross-Sections}: The Hu model approximated the NUV absorption cross-sections of CO$_2$ by its total extinction cross-sections out to 270 nm. However, measurements indicate that for $>201.58$ nm CO$_2$, extinction is scattering-dominated (\citealt{ityaksov2008co2}) even though the reported bond dissociation energy of $532.2\pm0.4$ kJ mol$^-1$ \citep{Darwent1970} corresponds to 225 nm. In high-CO$_2$ scenarios, this error shielded H$_2$O from NUV photolysis, regardless of assumptions regarding H$_2$O NUV absorption. This lead to underestimates of H$_2$O photolysis rates and [OH], and hence overestimates of [CO], [O], and [O$_2$]. We corrected the CO$_2$ cross-sections in the Hu model to correspond to absorption \citep{ityaksov2008co2, Keller-Rudek2013}.
    \item \textbf{Correction of Reaction Networks}: We identified several errors in the reaction networks of the ATMOS and Kasting models. In the ATMOS model, these errors lead SO$_2$ to suppress CO, so that $r_{CO}$ was low regardless of assumptions on NUV H$_2$O absorption. In the Kasting model, these errors did not affect the baseline scenario, but lead to SO$_2$ suppression of CO in low-outgassing planetary scenarios. These errors and their suggested corrections are summarized in Appendix~\ref{sec:detailed_model_intercomparison}, Table~\ref{tab:reaction_network_updates}.
    \item \textbf{Self-Consistent Temperature-Pressure Profile}: The temperature-pressure profile in the ATMOS \texttt{Archaean+haze} scenario features a warm stratosphere, due to shortwave stratospheric heating from high-altitude haze ultimately sourced from high biogenic CH$_4$ emission. This leads to wet stratosphere and high H$_2$O photolysis rates. However, on a world lacking vigorous CH$_4$ production (e.g., an abiotic world), [CH$_4$] is low, haze is not expected to form, and CO$_2$-rich anoxic atmospheres are expected to have had cold, dry stratospheres \citep{DeWitt2009, Guzman-Marmolejo2013, Arney2016}. Therefore, when employing ATMOS \texttt{Archaean+haze} to this planet scenario, it is necessary to first calculate a consistent temperature-pressure profile.  
\end{itemize}

We identified the following points of difference between our models:

\begin{itemize}
    \item \textbf{Binary Diffusion Coefficient $nD(X,Y)$}: Use of a generalized formulation (Equation~\ref{eqn:gen_dif}) to estimate $nD(X,Y)$, relevant to diffusion-limited atmospheric escape, overestimates $nD(H_2,N_2)$ and $nD(H_2,CO_2)$ relative to laboratory measurements \citep{Banks1973, Marrero1972}, and hence underestimates pH$_2$ and $r_{CO}$. Surprisingly, $nD(H,CO_2)$ has not yet been measured; we recommend this as a target for future laboratory studies.
    \item \textbf{$CO + OH$ Rate Law}: Prescriptions for the rate constant of the reaction $CO+OH\rightarrow CO_2+H$ have evolved significantly. Moving forward, we recommend the prescription of \citet{Burkholder2015}, which is the most up-to-date known, and is intermediate relative to the prescriptions incorporated into our models to date.
    \item \textbf{$CO + S + M$ Rate Law}: The reaction $CO + S + M \rightarrow OCS + M$ has not been measured in the laboratory, but has been invoked to explain Venusian OCS \citep{Krasnopolsky2007, Yung2009}. Assumption of this reaction modestly reduces $r_{CO}$. We identify this as a key reaction for laboratory follow-up; confirmation of this reaction mechanism will affirm our understanding of Venusian atmospheric chemistry, while refutation will signal a need to closely re-examine our photochemical models of Venus and Venusian exoplanets.
\end{itemize}

If we repair these errors and align these parameters between our models, we find that our model predictions of $r_{CO}$ agree to within a factor of $2$, and that we can reproduce both the low ($\sim$200 ppm) and high ($\sim8200$ ppm) estimates for $r_{CO}$ that have been reported in the literature (Section~\ref{sec:dance_models_dance}; Appendix~\ref{sec:detailed_model_intercomparison}, Table~\ref{tab:high_and_low_co}). The overwhelmingly dominant factor is the prescription adopted for the NUV absorption cross-sections of H$_2$O(g). Prescriptions that omit this absorption (e.g., \citealt{Sander2011}) lead to high $r_{CO}$, and prescriptions that include this absorption (e.g., \citealt{Kasting1981}) lead to low $r_{CO}$. The  absorption measured in this work is higher in the NUV than considered by any of these prescriptions (Figure~\ref{fig:new_h2o_xc_inmodels}), implying $r_{CO}$ to be lower than previously calculated by any model (Section~\ref{sec:updated_model}). We conclude that we have resolved the disagreements between our models as measured by predictions of $r_{CO}$.

\section{Updated Photochemical Model \label{sec:updated_model}}
We include our newly measured H$_2$O cross-sections in the corrected Hu model, using both the extrapolated and cutoff prescriptions detailed in Section~\ref{sec:new_h2o_theory_comp}. We verify that the model still reproduces the atmospheres of modern Earth and modern Mars as detailed in \citet{Hu2012}. Table~\ref{tab:new_results} presents the effects of the new H$_2$O cross-sections on $r_{CO}$ and $r_{O_2}$ for the CO$_2$-dominated exoplanet scenario of \citet{Hu2012}. For these calculations, we returned to the simulation parameters as originally prescribed by \citet{Hu2012}, following the rationale given therein and to facilitate comparison with past results. In other words, we followed the simulation parameters tabulated in Table~\ref{tab:detailed_parameters}, not the uniform parameters adopted for the model intercomparison in Appendix~\ref{sec:dance_models_dance}. For all model runs, we verified maintenance of atmospheric redox balance \citep{Hu2012}.


\begin{deluxetable}{lcccc}[h]
\tablecaption{Summary of results from corrected model (i.e. with corrected CO$_2$ cross-sections), with results from uncorrected model also shown for context. The first sub-table, titled ``Standard Scenario", illustrates the effect of different prescriptions for H$_2$O NUV absorption on CO and O$_2$ concentrations in our standard scenario (Table~\ref{tab:species_bcs}). The second sub-table, titled ``Reduced Outgassing Scenarios", illustrates the effect of our new cross-sections on the abiotic O$_2$ buildup scenario reported in \citet{Hu2012}. The enhanced H$_2$O photolysis efficiently recombines CO and O$_2$, and removes this abiotic false-positive scenario for O$_2$. In this table, $r_X$ is the surface mixing ratio of $X$ (relative to dry CO$_2$+N$_2$). $J_{H_{2}O}$ is the column-integrated photolysis rate. $\Phi_{Dep}$ is the net flux of reducing power out of the ocean, in H-equivalents, relevant to the question of global redox balance ($\Phi_{Dep}<0 \Rightarrow$ reducing power enters the ocean). \label{tab:new_results}}
\tablewidth{0pt}
\tablehead{
Parameters &\colhead{$r_{CO}$} &\colhead{$r_{O_{2}}$} &\colhead{$J_{H_{2}O}$} &\colhead{$\Phi_{Dep}$} \\
 & & &\colhead{(cm$^{-2}$ s$^{-1}$)} & \colhead{(H cm$^{-2}$ s$^{-1}$)}
 }
\startdata
\cutinhead{\textbf{Standard Scenario}}
Uncorrected model & 8.2E-3& 1.5E-14 & 1.0E8 & -4.2E9\\
\citet{Sander2011} H$_2$O  & 6.4E-3 & 9.2E-19 & 9.3E7 & -3.7E9\\
\citet{Kasting1981} H$_2$O   & 1.2E-4 & 9.5E-19 & 1.2E10 & -2.0E10\\
Cutoff H$_2$O (this work)  & 8.6E-6& 7.2E-12 & 6.6E10& -6.0E10\\
Extrapolated H$_2$O (this work)  & 1.3E-5 & 2.5E-12 & 5.3E10 &-5.8E10\\
\cutinhead{\textbf{Reduced Outgassing Scenarios}}
\sidehead{$\phi_{H_{2}}=3\times10^{9}$ cm$^{-2}$ s$^{-1}$}
Uncorrected model   & 1.2E-2 & 3.2E-6 & 3.0E7 &  -4.3E9\\
\citet{Sander2011} H$_2$O   & 9.5E-3 & 3.0E-6 & 2.9E7 &  -3.7E9\\
Extrapolated H$_2$O (this work)  & 1.5E-5 & 2.8E-11 & 5.3E10 & -8.5E9\\
\sidehead{$\phi_{H_{2}}=\phi_{CH_{4}}=0$ cm$^{-2}$ s$^{-1}$}
Uncorrected model & 3.7E-2 & 1.5E-3 & 9.1E6 &  -1.8E9\\
\citet{Sander2011} H$_2$O  & 1.6E-2 & 2.1E-4 & 9.8E6 &  -1.8E9\\
Extrapolated H$_2$O (this work)  & 1.6E-5 & 3.4E-11 & 5.3E10 & -4.8E8\\
\enddata
\end{deluxetable}

\subsection{Effect of New Cross-Sections on $r_{CO}$ and $r_{O_2}$}
\begin{figure}[h]
\plotone{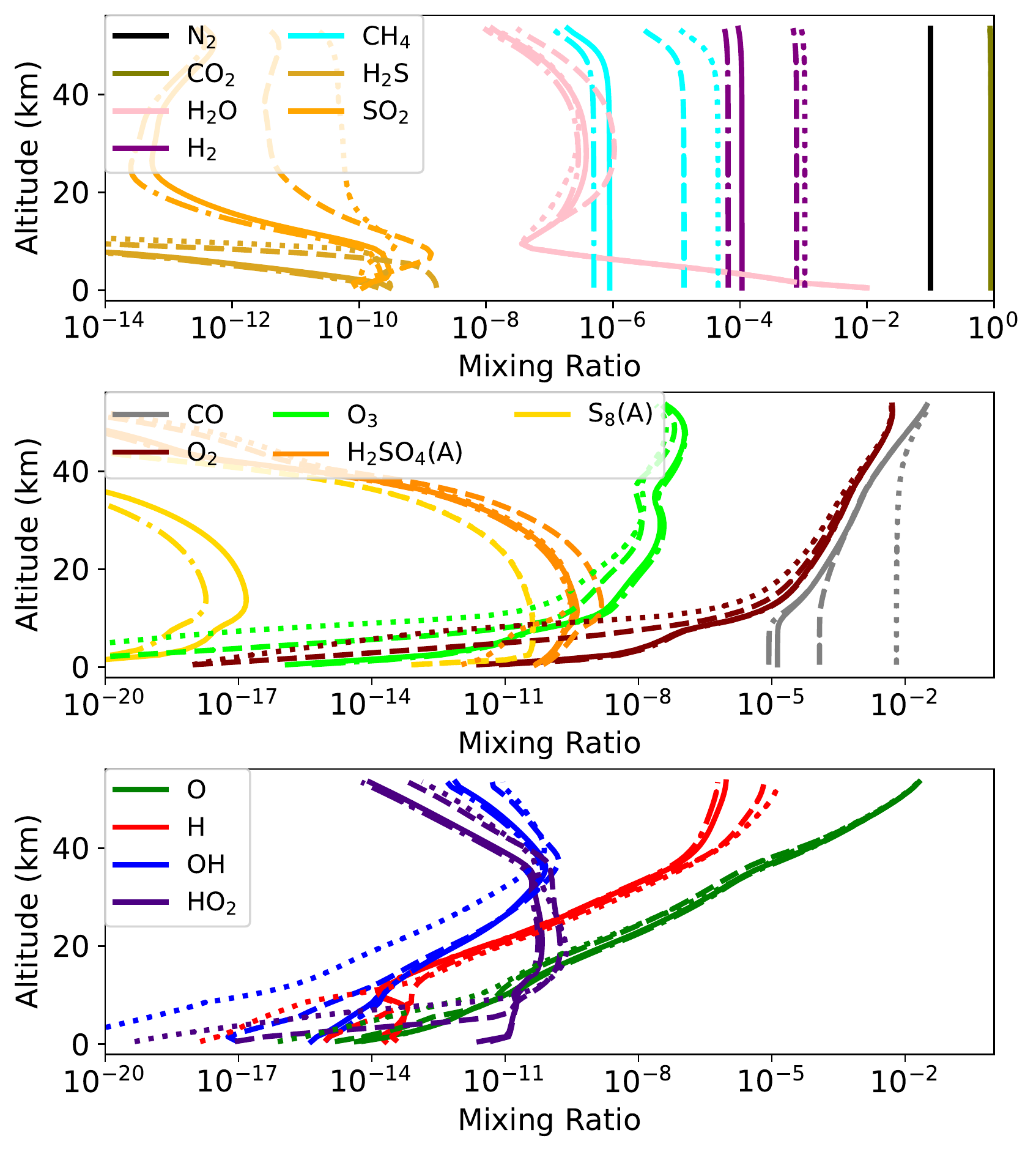}
\caption{Mixing ratio (relative to dry CO$_2$/N$_2$) as a function of altitude for outgassed species (top), photochemical byproducts (middle), and radicals (bottom). These predictions were derived from the Hu model, with corrected CO$_2$ cross-sections. The line types demarcate different prescriptions regarding the H$_2$O cross-sections. Specifically, the solid lines refer to the ``extrapolation" prescription proposed here; the dash-dotted lines, the ``cutoff" prescription; the dashed lines, the \citet{Kasting1981} prescription; and the dotted lines, the \citet{Sander2011} prescription. The higher H$_2$O absorption we propose here leads to higher [OH] and [H], and concommitantly lower [CO], [CH$_4$], and [H$_2$]. \label{fig:new_h2o_model_concs}}
\end{figure}

\begin{figure}[h]
\plotone{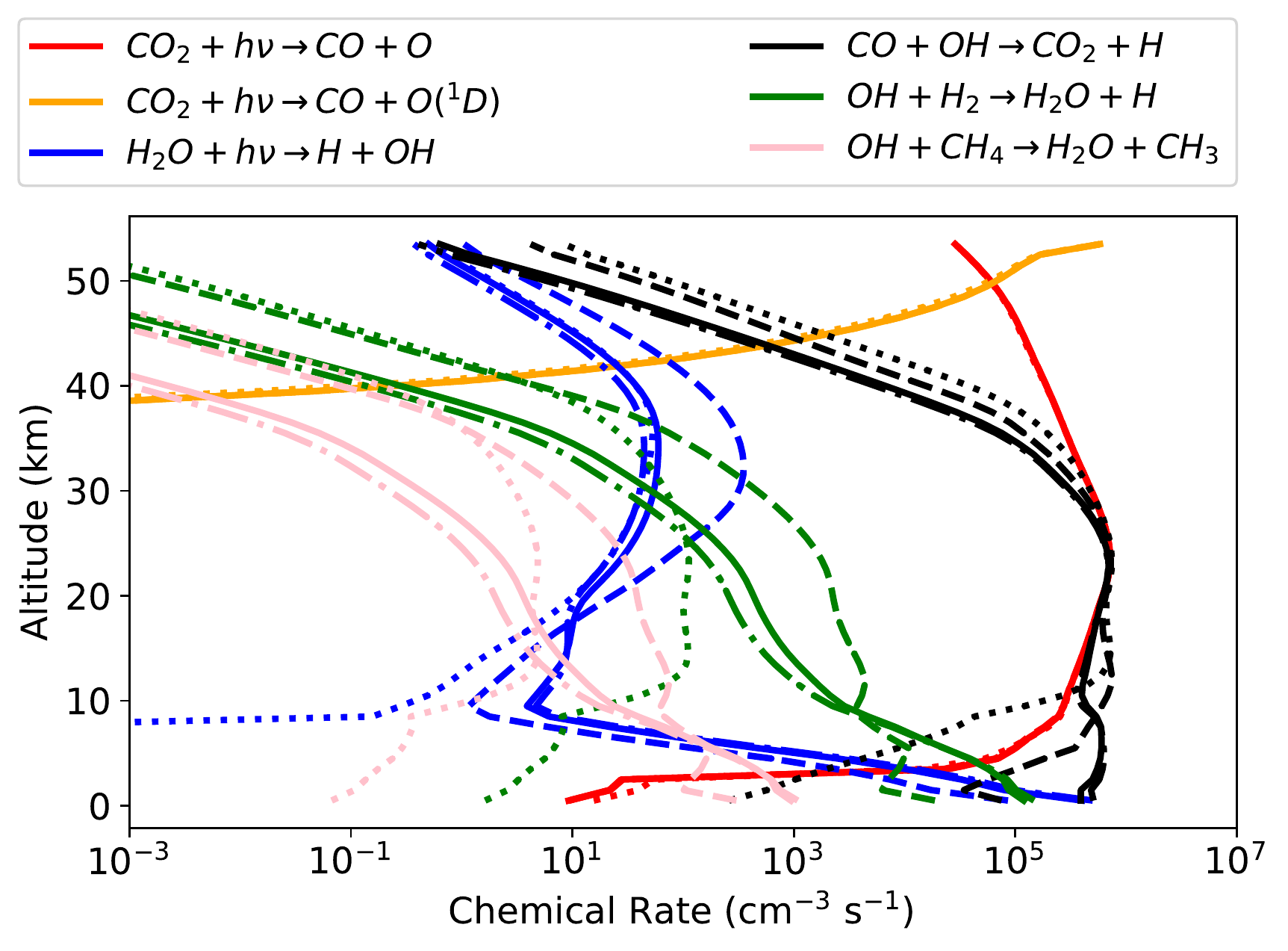}
\caption{Key reaction rates as a function of altitude. These predictions were derived from the Hu model, with corrected CO$_2$ cross-sections. The line types demarcate different prescriptions regarding the H$_2$O cross-sections. Specifically, the solid lines refer to the ``extrapolation" prescription proposed here; the dash-dotted lines, the ``cutoff" prescription; the dashed lines, the \citet{Kasting1981} prescription; and the dotted lines, the \citet{Sander2011} prescription. The higher H$_2$O absorption we propose here dramatically enhances OH and H production from H$_2$O photolysis, especially in the bottom of the atmosphere, leading to much lower CO, CH$_4$, and H$_2$ due to suppression by OH.\label{fig:new_h2o_model_rates}}
\end{figure}

The effects are dramatic: inclusion of the new H$_2$O cross-sections, whether using the extrapolated or cutoff prescriptions, reduces $r_{CO}$ by 2.5 orders of magnitude relative to the cross-sections recommended by \citet{Sander2011} (i.e., terminated at 198 nm), and 1 order of magnitude relative to the \citet{Kasting1981} prescription, to $\sim10$ ppm (Figure~\ref{fig:new_h2o_model_concs}), for our abiotic CO$_2$-N$_2$ scenario. The new H$_2$O cross-sections are larger and extend to longer wavelengths than the prescription of \citet{Kasting1981}, leading to H$_2$O photolysis rates that are $\sim5$ times higher. These higher photolysis rates drive enhanced production of OH, especially at the bottom of the atmosphere where [H$_2$O] is highest (Figure~\ref{fig:new_h2o_model_rates}), resulting in the efficient recombination of CO and O to CO$_2$ via catalytic cycles ultimately triggered by $CO+OH\rightarrow CO_2+H$ \citep{Harman2018}. The new cross-sections also drive enhanced production of H, visible as enhanced [H] in the bottom of the atmosphere.

\begin{figure}[h]
\plotone{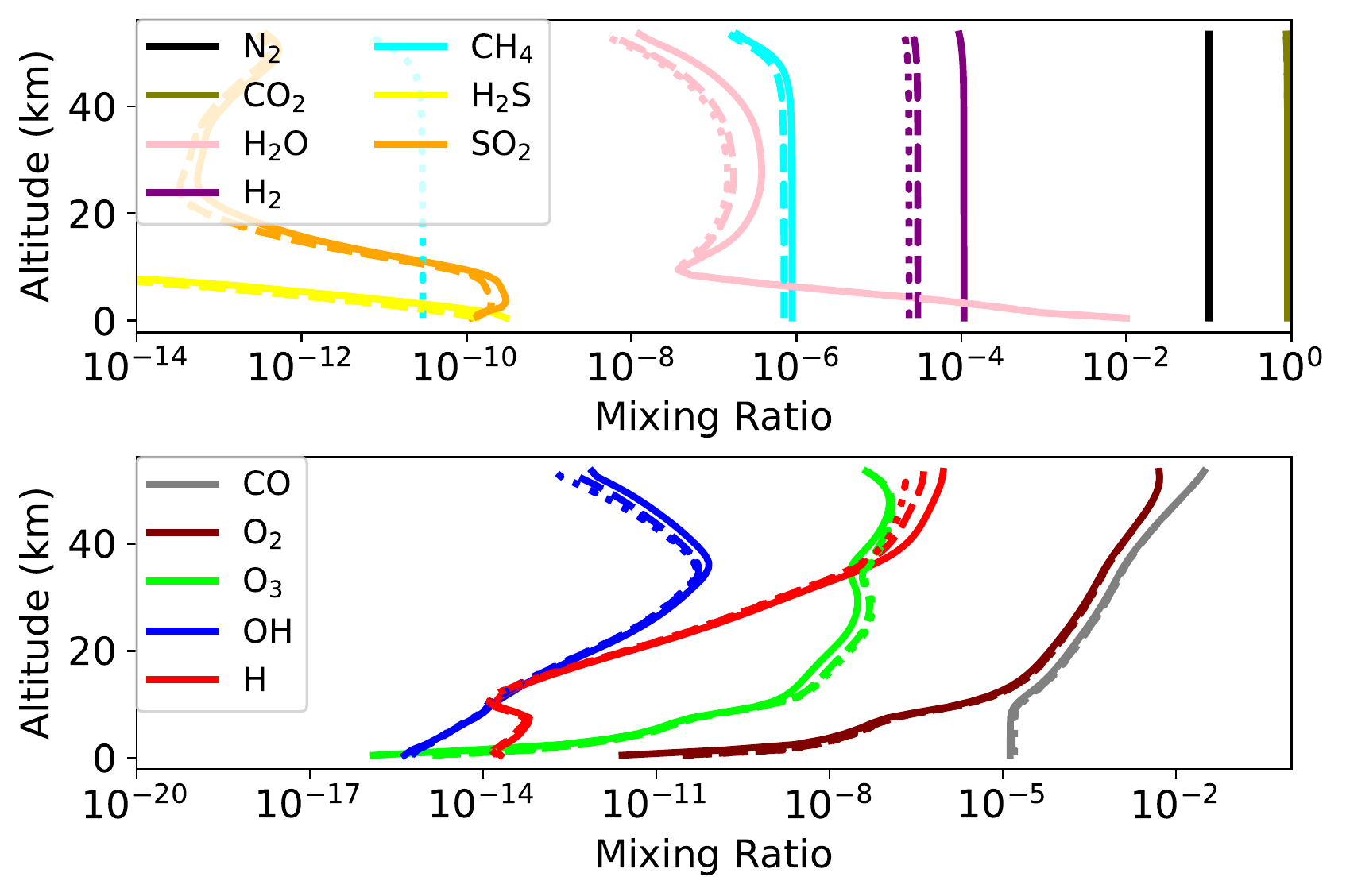}
\caption{Mixing ratio (relative to dry CO$_2$/N$_2$) as a function of altitude for outgassed species (top) and photochemical byproducts (bottom), calculated with the corrected Hu model and with the extrapolation prescription for H$_2$O absorption. The solid lines assume full outgassing ($\phi_{H_{2}}=3\times10^{10}$ cm$^{-2}$ s$^{-1}$, $\phi_{CH_{4}}=3\times10^{8}$ cm$^{-2}$ s$^{-1}$); the dashed lines, reduced outgassing ($\phi_{H_{2}}=3\times10^{9}$ cm$^{-2}$ s$^{-1}$,  $\phi_{CH_{4}}=3\times10^{8}$ cm$^{-2}$ s$^{-1}$); and the dotted lines no outgassing of H$_2$ and CH$_4$ ($\phi_{H_{2}}=0$ cm$^{-2}$ s$^{-1}$. The higher H$_2$O absorption we measure obviates the low-outgassing false photochemical false positive mechanism for O$_2$. \label{fig:new_h2o_model_concs_outgassing}}
\end{figure}

Our H$_2$O absorption cross-sections also negate the low-outgassing photochemical false positive scenario for O$_2$ on planets orbiting Sun-like stars. Specifically, it has been proposed that in the regime of lower outgassing of reductants (H$_2$, CH$_4$), O$_2$ sourced from CO$_2$ photolysis can accumulate to detectable, near-biotic levels on planets orbiting solar-type stars. This constitutes a potential false positive scenario for O$_2$ as a biosignature gas \citep{Hu2012, Harman2015, James2018}. With our new H$_2$O cross-sections, we find photolytic OH efficiently recombines CO and O even in the absence of CH$_4$ and H$_2$ outgassing (Figure~\ref{fig:new_h2o_model_concs_outgassing}). Interestingly, though we find very low H$_2$ and CH$_4$ in the low-outgassing case, we nonetheless report higher pH$_2$ and pCH$_4$ compared to \citet{Hu2012}. We speculate H sourced from H$_2$O photolysis to support the CH$_4$ and H$_2$.  

 \subsection{Effect of New Cross-Sections on Other Atmospheric Gases}
OH is a powerful oxidizing agent, and efficiently reacts with a broad range of reduced species (e.g., \citealt{CatlingKasting2017}). It is consequently unsurprising to find that the enhancement in OH production from our larger H$_2$O cross-sections leads to suppression of a broad range of the trace compounds present in anoxic atmospheres, including H$_2$S, SO$_2$, and S$_8$ aerosol (Figure~\ref{fig:new_h2o_model_concs})

Perhaps most dramatic is the suppression of CH$_4$. With our new H$_2$O cross-sections, we predict the concentration of volcanically-outgassed CH$_4$ to be 2 orders of magnitude lower than using the \citet{Sander2011} cross-sections and 1.5 orders of magnitude lower than using the \citet{Kasting1981} cross-sections. We predict the main sink on CH$_4$ to be OH via the reaction $OH+CH_4\rightarrow CH_3+H_2O$, consistent with previous work (e.g., \citealt{Rugheimer2018}). This suppression of CH$_4$ is significant because CH$_4$ is spectrally active, and has been proposed as a potentially detectable component of exoplanet atmospheres and a probe of planetary processes, including life \citep{Guzman-Marmolejo2013, Rugheimer2018, Krissansen-Totton2018}. Our work suggests it may be harder to detect this gas in anoxic atmospheres than previously considered. 

Also key is our finding of photochemical suppression of H$_2$ (but see caveat below). Prior simulations concluded that the main sink on H$_2$ on anoxic terrestrial planets (e.g. early Earth) was escape to space and that pH$_2$ was to first order set by the balance between H$_2$ outgassing and (diffusion-limited) escape \citep{Kasting1993, Kasting2014}. Indeed, under the assumption of the \citet{Kasting1981} cross-sections, we recover this result ourselves. However, with our new cross-sections, we find the sink due to the reaction $H_2+OH\rightarrow H_2O+H$ to be the dominant sink on H$_2$, which suppresses p$H_2$ by 1 order of magnitude relative to past predictions. This reaction converts relatively unreactive H$_2$ to relatively reactive H, which can undergo further reactions to ultimately be deposited to the surface in the form of more-soluble reduced chemical species. This is reflected in enhanced transfer of reductants from the atmosphere to the oceans calculated by our model (Table~\ref{tab:new_results}), in which essentially all of the reducing power outgassed as H$_2$ is returned to the surface, primarily via rainout. This suggests more efficient delivery of reduced organic compounds from the atmosphere, of relevance to origin-of-life studies (e.g., \citealt{Cleaves2008, Harman2013, Rimmer2019}).

\begin{figure}[h]
\plotone{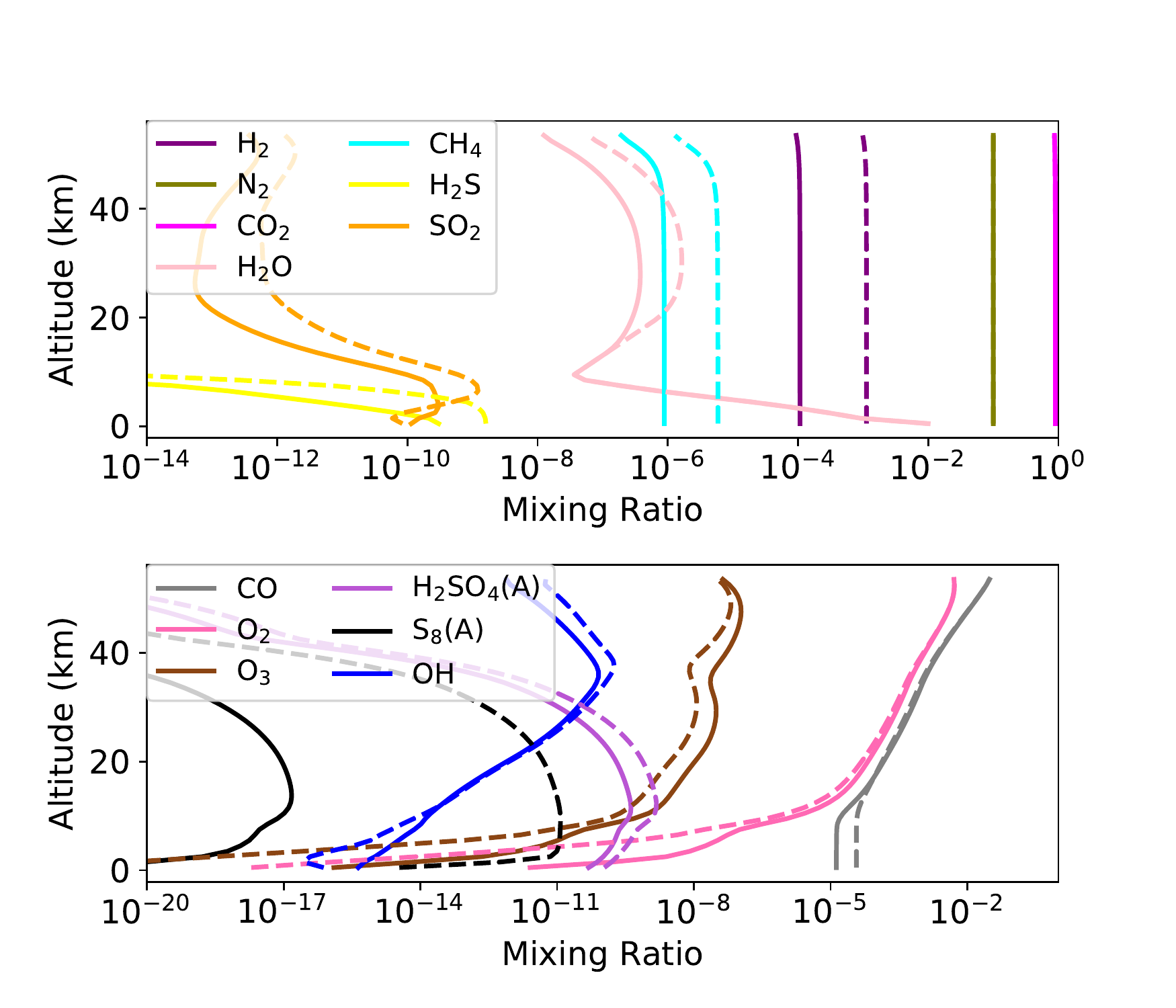}
\caption{Mixing ratio (relative to dry CO$_2$/N$_2$) as a function of altitude for outgassed species (top) and photochemical byproducts (bottom), calculated with the corrected Hu model and with the extrapolation prescription for H$_2$O absorption. The solid lines were calculated without assuming global redox balance; the dashed lines assumed global redox balance. Redox balance makes the atmosphere more reducing, stabilizing reduced species. \label{fig:redoxbalance}}
\end{figure}

The above results were derived without assuming global redox balance \citep{Kasting2013, Tian2014, Harman2015, James2018}. The principle of global redox balance is based on the observation that the main mechanisms by which we know free electrons to be added or removed from the ocean-atmosphere system on Earth are oxidative weathering and biologically-mediated burial of reductants. The former is not relevant to anoxic atmospheres; the latter is not relevant to abiotic worlds. If one zeroes these terms in the atmosphere-ocean redox budgets, one finds that in steady-state, any supply of reductants or oxidants to the planet surface should be counterbalanced by return flux to or deposition from the atmosphere \citep{Harman2015}. This is practically implemented in models by prescribing an oceanic H$_2$ return flux in the case where reductants are net deposited into the ocean by the atmosphere (mostly by rainout), or an increased H$_2$ deposition velocity in the (uncommon) case where oxidants are net deposited \citep{Tian2014, Harman2015, James2018}. Figure~\ref{fig:redoxbalance} presents the effects of requiring global redox balance. We find a return flux of $4.4\times10^{10}$ cm$^{-2}$ s$^{-1}$, a return of pH$_2$ to the escape-limited $1\times10^{-3}$ bar, and an increase in CH$_4$ by 1 order of magnitude ($2\times$ lower than when assuming the \citealt{Kasting1981} cross-sections). Overall, we predict pH$_2$ and pCH$_4$ to be significantly an order-of-magnitude higher on worlds obeying global redox balance, as has been proposed for abiotic worlds. 

Our calculations highlight the importance of the question of global redox balance for abiotic planets. In the past, it has been possible to largely ignore this question in conventional planetary scenarios, because the deposition terms in the redox budget have been relatively small. However, our new cross-sections suggest that the processing of H$_2$ into soluble reductants is efficient, the deposition terms are large, and the assumption of global redox balance has significant impact on the buildup of spectrally-active, potentially-detectable species in conventional planetary scenarios.

Whether abiotic anoxic planets are in global redox balance requires careful consideration. The theory of global redox balance rests on the assumption that biological mediation is required for burial of reductants. Biologically-mediated burial is the dominant mode on modern Earth \citep{Walker1974}. However, to our knowledge it is not yet determined whether biotic burial is the \emph{only} possible mode of reductant burial. We may draw an analogy to the theory of abiotic nitrogen fixation, where it was long assumed that on abiotic worlds (e.g, prebiotic Earth), lightning-fixed nitrogen could accumulate almost indefinitely in the ocean as nitrate/nitrite (NO$_X^-$), since the today-dominant biological sinks of NO$_X^-$ were absent (e.g., \citealt{Mancinelli1988, Wong2017, Hu2019}). However, there exist abiotic sinks on NO$_X^-$, slower than the biotic sinks but still important on geological timescales; these sinks suppress oceanic [NO$_X^-$] and stabilize atmospheric N$_2$ \citep{Ranjan2019}. Similarly, there may exist abiotic reductant/oxidant burial mechanisms which are relevant on geological timescales (e.g., Fe$^{2+}$ photooxidation, \citealt{Kasting1984}; magnetite burial, \citealt{James2018}); the existence of such mechanisms should be explored further. Alternately, Mars may provide a touchstone. Like Earth, early Mars should have hosted an abiotic ocean under an anoxic atmosphere, but unlike Earth, the lack of tectonic activity and hydrology means that geological evidence from this epoch may be preserved \citep{Citron2018, Sasselov2020}. Are there geological fingerprints of the presence or absence of redox balance (e.g, evidence of widespread abiotic reductant burial) that future missions might detect? If so, such measurements could bound the relevant parameter space \text{for} the global redox balance hypothesis.

\section{Discussion \& Conclusions \label{sec:disc_conc}}
We present the first measurements of H$_2$O cross-sections in the NUV ($>200$ nm) at habitable temperatures ($T=292~\text{K}<373$ K), and show them to be far higher than assumed by previous prescriptions. These cross-sections are critical because in anoxic atmospheres the atmosphere is transparent at these wavelengths, and water can efficiently photolyze down to the surface \citep{Harman2015, Ranjan2017a}. In anoxic atmospheres, this H$_2$O photolysis is the ultimate source of atmospheric OH, a key control on atmospheric chemistry in general and CO in particular.

To assess the photochemical impact of these new cross-sections on atmospheric composition, we apply a photochemical model to a planetary scenario corresponding to an abiotic habitable planet with an anoxic, CO$_2$-N$_2$ atmosphere orbiting a Sunlike star. This planet scenario is representative of early (prebiotic) Earth, Mars and Venus, and analogous exoplanets. Model predictions of the atmospheric composition of such worlds are highly divergent in the literature; through a model intercomparison, we have identified the errors and divergent assumptions driving these differences, and reconciled our models.

Incorporating these newly-measured cross-sections into our corrected model enhances OH production and suppresses $r_{CO}$ by $1-2.5$ orders of magnitude relative to past calculations. This implies less CO on early Earth for prebiotic chemistry and primitive ecosystems \citep{Kasting2014}, suggesting the need to consider alternate reductants. It also implies that CO will be a more challenging observational target for rocky exoplanet observations that we might previously have hoped. However, if surface production of CO from processes like impacts, volcanism or biology \citep{Kasting2014, Schwieterman2019, Wogan2020} is sufficient to saturate the enhanced OH sink due to more efficient H$_2$O photolysis, CO may yet enter runaway and build to potentially-detectable concentrations; we plan further investigation. On the other hand, the more efficient OH-catalyzed recombination of CO and O also removes the proposed low-outgassing false-positive mechanism for O$_2$ \citep{Hu2012}. This reduces (but does not completely obviate; \citealt{Wordsworth2014}) the potential ambiguities regarding O$_2$ as a biosignature for planets orbiting Sunlike stars. 

The situation on planets orbiting lower-mass stars, e.g. M-dwarfs, may be different. These cooler stars are dimmer in the NUV compared to Sunlike stars, meaning we expect the effect of our enhanced NUV H$_2$O cross-sections to be muted \citep{Segura2005, Ranjan2017c}. On these planets, higher $r_{CO}$ and $r_{O_{2}}$ may be possible \citep{Schwieterman2019, Hu2020}; we plan further study. Further, these results do not impact O$_2$-rich planets analogous to the modern Earth, since in these atmospheres direct H$_2$O photolysis is a minor contributor to OH production. 

In addition to CO, H$_2$O-derived OH can suppress a broad range of species in anoxic atmospheres. In particular, the larger H$_2$O cross-sections we measure in this work lead to substantial enhancements in OH attack on H$_2$ and CH$_4$, suppressing these gases in the abiotic scenario by 1-2 orders of magnitude relative to past calculations, and suggesting that spectroscopic detection of CH$_4$ on anoxic exoplanets will be substantially more challenging than previously considered \citep{Reinhard2017, Krissansen-Totton2018}. However, this finding is sensitive to assumption of global redox balance. If reductants cannot be removed from the ocean by burial, as has been proposed for abiotic planets, then the return flux of reductants from the ocean (parametrized as H$_2$) compensates for much of the CH$_4$ and all of the H$_2$ suppression. Regardless, rainout of reductants to the surface is enhanced, relevant to prebiotic chemistry (c.f. \citealt{Benner2019Thanos})

In prior calculations, enforcement of global redox balance resulted in relatively small changes in many planetary scenarios, including the scenario studied here. With the enhanced OH production driven by our higher H$_2$O cross-sections, this is no longer the case. This highlights the need to carefully consider global redox balance, and in particular its key premise that abiotic reductant burial is always geologically insignificant. Early Mars may provide a test case for the theory of redox balance, in that it may have hosted an abiotic ocean underlying an anoxic atmosphere early in its history, and geological remnants of this era might persist due to the lack of hydrologic and tectonic activity since 3.5 Ga. If abiotic reductant burial produce a detectable geological signature, then future missions can search for that signature, directly testing the global redox balance hypothesis. Further work, to consider processes and signatures of abiotic reductant burial, is required.

In this paper, we have focused on an abiotic planet scenario. We note that all of our specific findings (e.g., very low $r_{CO}$) may not generalize to biotic scenarios. Biological production or uptake of gases may significantly outpace photochemical sources and sinks; for example, if biological CO production can outpace photolytic OH supply, then CO may nonetheless build to high, potentially-detectable concentrations \citep{Schwieterman2019}. Detailed case-by-case modelling of biotic scenarios is required. However, our general point that OH production is higher than previously considered on anoxic habitable planets applies to biotic worlds as well, implying that spectrally active trace gases have a higher bar to clear to build to high concentrations than previously considered.

In this work, we have ignored nitrogenous chemistry, in particular the NO$_X$ catalytic chemistry triggered by lightning-generated NO \citep{Ardaseva2017, Harman2018}. We justify this exclusion on the basis that this chemistry is most important when CO is high \citep{Kasting1990}, and our models indicate that CO is photochemically suppressed. We conducted a sensitivity test to the inclusion of NO-triggered nitrogenous chemistry with the Kasting model, and found negligible (percent-level) impact on $r_{CO}$. Note that nitrogenous chemistry has been proposed to play a more dominant role on M-dwarf planets \citep{Hu2020}; for such worlds, this chemistry must be included.

Our work highlights the critical need for laboratory measurements and/or theoretical calculations of the inputs to photochemical models. We show the sensitivity of the models to H$_2$O NUV cross-sections; we recommend further characterization of these cross-sections, both to confirm our own results and to extend these cross-sections, e.g. to longer wavelengths and lower temperatures. In particular, we reiterate that our prescriptions for H$_2$O NUV cross-sections are conservative, and the true absorption may be yet higher; higher signal-to-noise measurements at longer wavelengths are required to rule on this possibility. Further, our 292K cross-sections are good proxies for H$_2$O absorption on temperate terrestrial planets, because the nonlinear decrease in H$_2$O saturation pressure with temperature means that most H$_2$O is confined to the temperate lower atmosphere. However, on cold planets (e.g., modern Mars), the lower atmosphere is also cold, meaning use of 292K cross-sections may overestimate the H$_2$O opacity and photolysis rate\footnote{At lower temperatures fewer energy levels can be populated, which decreases the total number of active transition frequencies and subsequent cross-sectional opacity (see, for example, \cite{schulz2002ultraviolet}).}. Similarly, we have here assumed a photolysis quantum efficiency of unity, i.e. that absorption of each $\leq230$ nm photon leads to H$_2$O photolysis. If this assumption is incorrect, then the true photolysis rate will be lower than we have modelled here. 

Finally, some of the reactions encoded into our models and/or their reaction rate constants are assumed or disputed; these reactions should be experimentally or theoretically characterized, to confirm or refute these assumptions. In particular, we identify the reactions $CO + S + M \rightarrow OCS + M$, $HO_2+SO_2\rightarrow OH+SO_3$, $HO_2+SO_2 \rightarrow O_2 + HSO_2$, $SO + HO_2 \rightarrow SO_2 + OH$, $SO+HO_2\rightarrow SO_2+OH$, $SO+HO_2\rightarrow O_2+HSO$, and $N+O_3\rightarrow NO + O_2$ as targets for further investigation \citep{Graham1979,Yung1982, Wang2005, Yung2009, Burkholder2015}.

\acknowledgments

We thank Iouli Gordon, Eamon Conway, Robert Hargreaves, Mike Wong, Kevin Zahnle, Timothy Lee, Shawn Domagal-Goldman, Mark Claire, David Catling, Nick Wogan, and Jim Kasting for helpful discussions and answers to questions. We thank an anonymous reviewer for critical feedback which improved this work. This work was supported in part by a grant from the Simons Foundation (SCOL grant 495062 to S. R.) and the Heising-Simons Foundation (51 Pegasi b Fellowship to C.SS). E.W.S. gratefully acknowledges support by NASA Exobiology grant 18-EXO18-0005 and NASA Astrobiology Program grants NNA15BB03A and 80NSSC18K0829. The work partly has received funding from the EMPIR programme co-financed by the Participating States and from the European Union’s Horizon 2020 research and innovation programme (Grant Number 16ENV08). This research has made use of NASA's Astrophysics Data System, and the MPI-Mainz UV-VIS Spectral Atlas of Gaseous Molecules \citep{Keller-Rudek2013}.

%

\vspace{5mm}





\appendix

\section{Detailed Simulation Parameters for Planetary Scenario\label{sec:detailed_simulation_parameters}}
In Table~\ref{tab:detailed_parameters}, we present the simulation parameters of our models for the CO$_2$-dominated planet scenario we consider. To make our models agree, we must implement the corrections summarized in Table~\ref{tab:corrections}, and adjusting our models to use common inputs and formalisms as summarized in Table~\ref{tab:constant_parameters}.

\begin{deluxetable}{cccc}[h]
\tablecaption{Simulation Parameters For Planetary Scenario.\label{tab:detailed_parameters}}
\tablewidth{0pt}
\tablehead{Scenario Parameter & \colhead{Hu} & \colhead{ATMOS} & \colhead{Kasting}}
\startdata
Model & \citet{Hu2012} & \citet{Arney2016}  & \citet{Harman2015}\\
Reaction Network & As in \citet{Hu2012} & Archean Scenario   & As in \citet{Harman2015}\\
& (Excludes N-, C$_{>2}$-chem) & & \\
\hline
Stellar type & Sun & Sun & Sun \\ 
Semi-major axis & & 1.3 AU & \\
Planet size & & 1 M$_{\earth}$, 1 R$_{\earth}$ &  \\
Surface albedo &  0. & 0.25 & 0.25\\ 
Major atmospheric components & & 0.9 bar CO$_2$, 0.1 bar\ N$_2^\star$&  \\
Surface temperature ($z=0$ km) &  & 288K &  \\
Surface $r_{H_{2}O}$ (lowest atmospheric bin) &  & 0.01 &  \\
Eddy Diffusion Profile &  & See Figure~\ref{fig:tph2oeddy} &  \\
Temperature-Pressure Profile &  & See Figure~\ref{fig:tph2oeddy} &  \\
Vertical Resolution & 0-54 km, 1 km steps & 0-100 km, 0.5 km steps & 0-100 km, 1 km steps\\
\hline
Rainout & Earthlike; rainout turned off for   & Earthlike (all species) & Earthlike (all species) \\
& H$_2$, CO, CH$_4$, NH$_3$, N$_2$, C$_2$H$_2$, & & \\
& C$_2$H$_4$, C$_2$H$_6$, and O$_2$ to simulate & & \\
& saturated ocean on abiotic planet& & \\
Lightning & & Off & \\
Global Redox Conservation & No & No & Yes\\
\enddata
\tablecomments{$^\star$In Hu model, pN$_2$ is fixed. In the ATMOS and Kasting models, pN$_2$ is adjusted to maintain dry $P=1$ bar. Photochemical and outgassed products do not build to levels comparable to the N$_2$ inventory in the scenario simulated here, and consequently this difference does not affect our results}
\end{deluxetable}

\section{Detailed Boundary Conditions for Planetary Scenario \label{sec:detailed_boundary_conditions}}
In this Appendix, we present the species used in each of our models and their corresponding boundary conditions used in our initial model reconciliation. For all species, we assign either a fixed surface mixing ratio or a surface flux. CO$_2$, N$_2$, and H$_2$O are the only species assigned fixed surface mixing ratios (for rationale, see \citealt{Hu2012}). For species with a surface flux, the species is assumed to be injected in the bottommost layer of the atmosphere, i.e. PARAMNAME=1 in ATMOS. H and H$_2$ are assumed to escape at their diffusion-limited rates; for all other species, the escape/delivery flux is prescribed as 0. 

While our models generally assume the same major species and many of the same minor species, there are some key differences, driven primarily by different assumptions regarding reaction network. In particular:
\begin{enumerate}
    \item The Kasting model does not include polysulfur species. This is because the polymerization of elemental sulfur to form S$_8$ aerosol is ignored, on the basis that this is a minor exit for S in this relatively oxidized atmospheric scenario; instead, S is assigned a high deposition velocity of 1 cm~s$^{-1}$. 
    \item The Hu model excludes all nitrogenous species other than N$_2$. This is because the chosen boundary conditions  precluded reactive N (no lightning, no thermospheric N), meaning they could neglect N-chemistry. 
\end{enumerate}

In the planetary scenario considered here, these differences do not significantly affect $r_{CO}$, and we ignore them for purposes of this model intercomparison.

\startlongtable
\begin{deluxetable}{cccccccc}
\tablecaption{Species choice and treatment for the photochemical models used in this study.\label{tab:species_bcs}}
\tablewidth{0pt}
\tablehead{
\colhead{Species} &  & \colhead{Type} &  & \colhead{Surface Flux} & \colhead{Surface Mixing Ratio} & \colhead{Dry Deposition Velocity} & \colhead{TOA Flux}\\
  & \colhead{Hu}& \colhead{ATMOS} & \colhead{Kasting} & \colhead{(cm$^{-2}$ s$^{-1}$)} & \emph{(relative to CO$_2$+N$_2$)} & \colhead{(cm s$^{-1}$)} & \colhead{(cm$^{-2}$ s$^{-1}$)}
 }
\startdata
H & X & X & X & 0 & -- & 1 & Diffusion-limited \\
H$_2$ & X & X & X & $3\times10^{10}$ & -- & 0 & Diffusion-limited  \\
O & X & X & X & 0 & -- & 1 & 0 \\
O(1D) & X & F & F & 0 & -- & 0 & 0 \\
O$_2$ & X & X & X & 0 & -- & 0 & 0 \\
O$_3$ & X & X & F & 0 & -- & 0.4 & 0 \\
OH & X & X & X & 0 & -- & 1 & 0 \\
HO$_2$ & X & X & X & 0 & -- & 1 & 0 \\
H$_2$O & X & X & X & -- & 0.01 & 0 & 0 \\
H$_2$O$_2$ & X & X & X & 0 & -- & 0.5 & 0 \\
CO$_2$ & X & C & X & -- & 0.9 & 0 & 0 \\
CO & X & X & X & 0 & -- & $1\times10^{-8}$ & 0 \\
CH$_2$O & X & - & X & 0 & -- & 0.1 & 0 \\
CHO & X & - & X & 0 & -- & 0.1 & 0 \\
C & X & - & - & 0 & -- & 0 & 0 \\
CH & X & X & - & 0 & -- & 0 & 0 \\
CH$_2$ & X & - & - & 0 & -- & 0 & 0 \\
${}^{1}$CH$_2$ & X & F & F & 0 & -- & 0 & 0 \\
${}^{3}$CH$_2$ & X & X & F & 0 & -- & 0 & 0 \\
CH$_3$ & X & X & X & 0 & -- & 0 & 0 \\
CH$_4$ & X & X & X & $3\times10^{8}$ & -- & 0 & 0 \\
CH$_3$O & X & X & F & 0 & -- & 0.1 & 0 \\
CH$_4$O & X & - & - & 0 & -- & 0.1 & 0 \\
CHO$_2$ & X & - & - & 0 & -- & 0.1 & 0 \\
CH$_2$O$_2$ & X & - & - & 0 & -- & 0.1 & 0 \\
CH$_3$O$_2$ & X & X & F & 0 & -- & 0 & 0 \\
CH$_4$O$_2$ & X & - & - & 0 & -- & 0.1 & 0 \\
C$_2$ & X & X & - & 0 & -- & 0 & 0 \\
C$_2$H & X & X & - & 0 & -- & 0 & 0 \\
C$_2$H$_2$ & X & X & - & 0 & -- & 0 & 0 \\
C$_2$H$_3$ & X & X & - & 0 & -- & 0 & 0 \\
C$_2$H$_4$ & X & X & - & 0 & -- & 0 & 0 \\
C$_2$H$_5$ & X & X & F & 0 & -- & 0 & 0 \\
C$_2$H$_6$ & X & X & X & 0 & -- & $1\times10^{-5}$ & 0 \\
C$_2$HO & X & - & - & 0 & -- & 0 & 0 \\
C$_2$H$_2$O & X & - & - & 0 & -- & 0.1 & 0 \\
C$_2$H$_3$O & X & - & F & 0 & -- & 0.1 & 0 \\
C$_2$H$_4$O & X & - & - & 0 & -- & 0.1 & 0 \\
C$_2$H$_5$O & X & - & - & 0 & -- & 0.1 & 0 \\
S & X & X & - & 0 & -- & 0 & 0 \\
S$_2$ & X & X & - & 0 & -- & 0 & 0 \\
S$_3$ & X & F & - & 0 & -- & 0 & 0 \\
S$_4$ & X & F & - & 0 & -- & 0 & 0 \\
SO & X & X & X & 0 & -- & 0 & 0 \\
SO$_2$ & X & X & X & $3\times10^9$ & -- & 1 & 0 \\
${}^1$SO$_2$ & X & F & F & 0 & -- & 0 & 0 \\
${}^3$SO$_2$ & X & F & F & 0 & -- & 0 & 0 \\
H$_2$S & X & X & X & $3\times10^{8}$ & -- & 0.015 & 0 \\
HS & X & X & X & 0 & -- & 0 & 0 \\
HSO & X & X & X & 0 & -- & 0 & 0 \\
HSO$_2$ & X & - & - & 0 & -- & 0 & 0 \\
HSO$_3$ & X & F & F & 0 & -- & 0.1 & 0 \\
HSO$_4$ & X & X & X & 0 & -- & 1 & 0 \\
H$_2$SO$_4$(A) & A & A & A & 0 & -- & 0.2 & 0 \\
S$_8$ & X & - & - & 0 & -- & 0 & 0 \\
S$_8$(A) & A & A & - & 0 & -- & 0.2 & 0 \\
N$_2$ & C & C & C & -- & 0.1 & 0 & 0 \\
OCS	& X	& X & - & 0 & -- & 0.01 &	0 \\
CS	& X	& X & - &  0 & -- & 0.01 &	0 \\
CH$_3$S & X	& - & - & 0 & -- & 0.01 &	0 \\
CH$_4$S & X	& - & - & 0 & -- & 0.01 &	0 \\
\enddata
\tablecomments{(1) For species type, for each model, ``X" means the full continuity-diffusion equation is solved for the species; ``F" means it is treated as being in photochemical equilibrium; ``A" means it is an aerosol and falls out of the atmosphere; ``C" means it is treated as chemically inert; and ``--" means that it is not included in that model. Note that boundary conditions like dry deposition velocity are not relevant for Type ``F" species, since transport is not included for such species. The exclusion of a species from a model does not necessarily mean that the model is incapable of simulating the species, but just that it was not included in the atmospheric scenario selected here. For example, following \citet{Hu2012}, the Hu model was deployed here without N species because the planet scenario selected here precludes reactive N, though it is capable of simulating nitrogenous chemistry. (2) For the bottom boundary condition, either a surface flux is specified, or a surface mixing ratio. N$_2$ is a special case in the Kasting model and in ATMOS; in these models, [N$_2$] is adjusted to set the total dry pressure of the atmosphere to be 1 bar (to account for outgassed species and photochemical intermediates). Consequently, pN$_2\lesssim0.1$ bar in these models. (3) TOA flux refers to the magnitude of outflow at the top-of-the-atmosphere (TOA); hence, a negative number would correspond to an inflow.} 
\end{deluxetable}

\section{Detailed Model Intercomparison\label{sec:detailed_model_intercomparison}}
In this Appendix, we enumerate the model differences which drove the divergent predictions of $r_{CO}$. Some of these differences were errors; we discuss them below, and summarize their correction in Table~\ref{tab:corrections}.

\begin{deluxetable}{cc}[h]
\tablecaption{Required Corrections To the Models\label{tab:corrections}}
\tablewidth{0pt}
\tablehead{Model & Correction}
\startdata
Hu & Removal of erroneous $>202$ nm CO$_2$ absorption (Section~\ref{sec:xsec-co2-h2o}).\\
ATMOS, Kasting & Correction of reaction network (Section~\ref{sec:reac-network}) \\
ATMOS & Use of self-consistent T-P-H$_2$O Profile (Section~\ref{sec:atm_prof})
\enddata
\end{deluxetable}

\subsection{CO$_2$ and H$_2$O Cross-Sections\label{sec:xsec-co2-h2o}}
Our model intercomparison reveals the single strongest control on $r_{CO}$ to be treatment of the H$_2$O and CO$_2$ cross-sections. The abundance of CO in habitable planet atmospheres is photochemically controlled by a balance of CO$_2$ photolysis, which generates CO, and H$_2$O photolysis, which is the ultimate source of the OH radicals that recombine CO and O to CO$_2$ (c.f. Section~\ref{sec:CO_background}). The Hu model incorrectly implemented CO$_2$ absorption. Specifically, \citet{Hu2012} approximated the absorption cross-sections by the total-extinction cross-sections from \citet{ityaksov2008co2}. However, at wavelengths $>201.58$ nm, the extinction cross-section is dominated by scattering, and the absorption is $\approx 0$. Therefore, this error led to unphysical absorption from CO$_2$ and suppression of the UV radiation field at $\geq202$ nm. This suppression was particularly acute because of the high scattering optical depth at $\sim$200 nm in the CO$_2$-dominated atmosphere, which amplified the unphysical absorption due to CO$_2$ \citep{Bohren1987, Ranjan2017b}. CO$_2$ absorption at $\geq 204$ nm was not assumed to lead to photolysis, so CO$_2$ photolysis itself was not overestimated due to this error.

Upon correction of the CO$_2$ absorption cross-sections, the atmosphere is largely transparent at wavelengths $\geq$ 202 nm except for H$_2$O. Prior to this work, no experimentally measured or theoretically predicted absorption cross-sections were available for H$_2$O(g) at $>198$ nm at conditions relevant to temperate rocky planets ($T\sim300$ K) The cross-sections recommended by \citet{Burkholder2015}, ultimately sourced from \citet{Parkinson2003}, terminate at 198 nm due to lack of data. However, as originally pointed out by \citet{Kasting1981}, it is unphysical to assume that H$_2$O absorption should abruptly terminate at $\sim 200$ nm. The dissociation energy of the H-OH bond corresponds to photons of wavelength $\sim240$ nm, and photolysis should continue down to that wavelength, albeit at steadily lower cross-section. Consequently, models descended from \citet{Kasting1981}, in this study ATMOS and Kasting, extrapolate the H$_2$O cross-sections from \citet{Thompson1963} out to longer wavelengths (Figure~\ref{fig:new_h2o_xc_inmodels}). This leads to H$_2$O photolysis and CO recombination at low altitudes \citep{Harman2015}. The Hu model originally used this extrapolation as well; however, due to the error in the CO$_2$ cross-sections, H$_2$O photolysis was suppressed at low altitudes, regardless of whether this extrapolation was included, and the Hu model eventually adopted the recommendation of \citet{Burkholder2015} for the H$_2$O cross-sections. 

\begin{figure}[h] 
\plotone{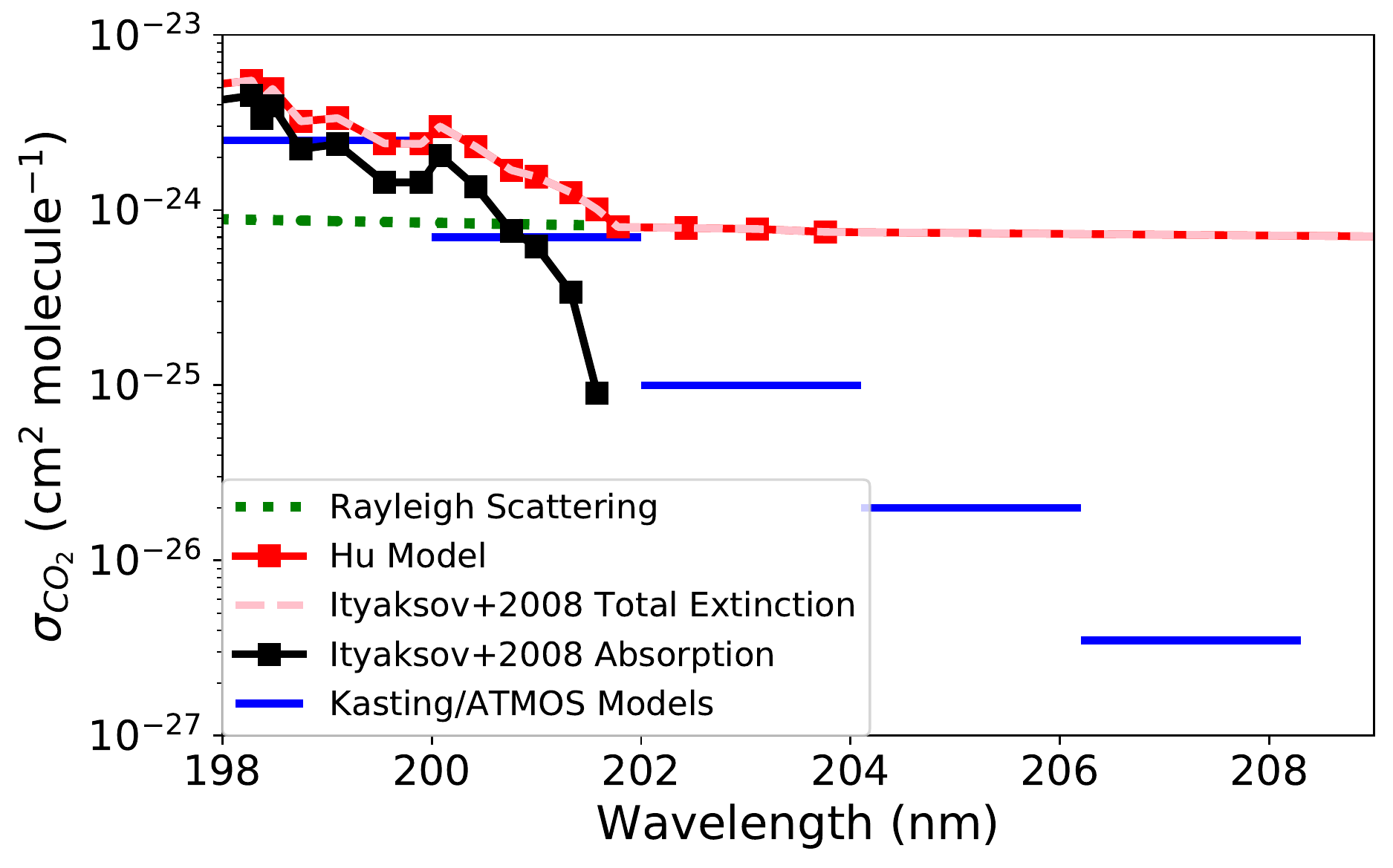}
\caption{CO$_2$ absorption cross-sections assumed by Hu and ATMOS/Kasting models. The total extinction and absorption cross-sections from \citet{ityaksov2008co2} are also shown for reference, as is the CO$_2$ Rayleigh scattering cross-section calculated as in \citet{Ranjan2017a}.\label{fig:co2-xc}}
\end{figure}

If \emph{either} the erroneous longwave CO$_2$ absorption is present, \emph{or} the H$_2$O absorption extrapolation is neglected, then H$_2$O photolysis is suppressed and $r_{CO}$ is high. CO$_2$ does not absorb  $>202$ nm, and in Section~\ref{sec:measurements} we experimentally demonstrate that H$_2$O does indeed absorb at such wavelengths. Indeed, it absorbs larger cross-sections than assumed by the ATMOS and Kasting models, meaning that H$_2$O photolysis, and hence CO recombination, is more intense than predicted by all three of the baseline models (Figure~\ref{fig:new_h2o_xc_inmodels}). Further, the latest data suggest CO$_2$ absorption terminates by 202 nm, shorter than all three of the models considered here,  meaning that CO$_2$ photolysis (and hence CO production) is lower than assumed by all three of the models (Figure~\ref{fig:co2-xc}). Therefore, CO should not only be lower than predicted by the baseline Hu model, it should be lower than predicted by all three of our models (Section~\ref{sec:updated_model}). 

\subsection{Reaction Network\label{sec:reac-network}}
In our intercomparison, we found SO$_2$ outgassing to suppress $r_{CO}$ in ATMOS. SO$_2$ did not suppress $r_{CO}$ in the Kasting model in the baseline scenario, but did in the Kasting model in the low-outgassing regime (0 CH$_4$, H$_2$ outgassing; \citealt{Hu2012}). We intercompared the Hu, Kasting, and ATMOS reaction networks, with particular emphasis on the sulfur chemistry. We identified a number of discrepancies in the Kasting and ATMOS models, summarized in Table~\ref{tab:reaction_network_updates}. The discrepancies in the Kasting models were incorrect implementations of published reactions or rates. The discrepancies in ATMOS were the deactivation of known reactions; the rationale for these deactivations is not known. The primary effect of the correction of these discrepancies is to remove the effect whereby SO$_2$ outgassing suppresses $r_{CO}$, in all regimes evaluated in this study. 

\begin{rotatetable}
\begin{deluxetable}{p{4 cm}p{3.8 cm}p{3.8 cm}p{5 cm}}
\tablecaption{Required updates to reaction networks \label{tab:reaction_network_updates}}
\tablewidth{0pt}
\tablehead{\colhead{Reaction} &\colhead{Old Parameter} & \colhead{Corrected Parameter} & \colhead{Reference}}
\startdata
\textbf{Kasting} & \nodata & \nodata& \nodata\\
SO$_2$ + O + M $\rightarrow$ SO$_3$ + M & $k_0=$~1.8E-33$(T/300 K)^{-2}$; $k_\infty=$4.2E-14$(T/300 K)^{-1.8}$ & $k_0=$~1.8E-33$(T/300 K)^{2}$; $k_\infty=$4.2E-14$(T/300 K)^{1.8}$ & \citet{Sander2011}\\ 
SO+O$_3\rightarrow$SO$_2$ + O$_2$ & k=3.6E-12$\exp(-1100/T)$ & k=3.4E-12$\exp(-1100/T)$ & \citet{Sander2011} \\ 
HS + HO$_2$ & $\rightarrow$ H$_2$S + O$_2$ & $\rightarrow$ HSO + OH & \citet{Stachnik1987} \\ 
S+CO$_2\rightarrow$ SO + CO & k=1.2E-11 & k=1.0E-20 & \citet{Yung1982}\\
\textbf{ATMOS} & \nodata& \nodata& \nodata\\
$^3$CH$_2$+H$_2\rightarrow$ CH$_3$+H & k=0 & k=5.0E-14 & \citet{Harman2015}\\ 
$^3$CH$_2$+CH$_4\rightarrow$ CH$_3$+CH$_3$ & k=0 & k=7.1E-12$\exp(-5051/T)$ & \citet{Harman2015}\\ 
SO+HO$_2$ $\rightarrow$ SO$_2$+OH & k=0 & k=2.8E-11 &\citep{Harman2015}\\ 
HSO$_3$+OH $\rightarrow$ H$_2$O+SO$_3$ & k=0 & k=1.0E-11 & \citet{Harman2015} \\ 
HSO$_3$+H $\rightarrow$ H$_2$+SO$_3$ & k=0 & k=1.0E-11 & \citet{Harman2015} \\ 
HSO$_3$+O$\rightarrow$ OH+SO$_3$ & k=0 & k=1.0E-11 & \citet{Harman2015} \\ 
HS+O$_2$ $\rightarrow$ OH+SO & k=0 & k=4.0E-19 & \citet{Harman2015} \\ 
HS+H$_2CO\rightarrow$ H$_2$S+HCO & k=0 & k=1.7E-11$\exp(-800/T)$ & \citet{Harman2015} \\ 
SO$_2$+HO$_2\rightarrow$ SO$_3$S+OH & k=0 & k=1.0E-18 & \citet{Harman2015} \\ 
S+CO$_2\rightarrow$ SO+CO & k=0 & k=1.0E-20 & \citet{Yung1982} \\ 
SO+HO$_2\rightarrow$ HSO+O$_2$ & k=0 & k=2.8E-11 & \citet{Harman2015} \\ 
HSO+NO$_2\rightarrow$ HNO+SO & k=0 & k=1.0E-15 & \citet{Harman2015} \\ 
\enddata
\end{deluxetable}
\end{rotatetable}

We note there persist differences between the reaction networks encoded in our models. \citet{Harman2015} encode the reaction HO$_2$ + SO$_2\rightarrow$OH + SO$_3$, following \citet{Graham1979}. However, formally \citet{Graham1979} report a nondetection of this reaction and an upper limit for the reaction rate. \citet{Hu2012} instead encode for the same reactants the reaction HO$_2$ + SO$_2\rightarrow$O$_2$ + HSO$_2$, following theoretical calculations by \cite{Wang2005}. Furthermore, \citet{Harman2015} includes the reactions SO + HO$_2\rightarrow$ SO$_2$ + OH and SO + HO$_2\rightarrow$ O$_2$ + HSO, following \citet{DeMore1992}. SO + HO$_2\rightarrow$ SO$_2$ + OH was proposed by \citet{Yung1982} in analogy to SO + ClO. These reactions are not recommended in later versions of the JPL Evaluations \citep{Sander2011}; consequently, \citet{Hu2012} omit them.  Similarly, \citet{Harman2015} include the reaction N+O$_3\rightarrow$ NO+O$_2$, but ATMOS does not include this reaction, following the recommendation of \citet{Burkholder2015}. These differences in reaction network did not affect the results in the scenarios studied in this paper, but may be relevant to future work. 

\subsubsection{CO + OH Ratelaw}
Hu, ATMOS, and Kasting encode different rate laws for the reaction of CO and OH. Hu encodes it as a two-body reaction, with rate law (\citealt{Baulch1992} via NIST):

\begin{eqnarray}
    CO + OH \rightarrow CO_2 + H\nonumber \\
    k=5.4\times10^{-14}(\frac{T}{298~K})^{1.5} \exp(\frac{250.0~K}{T}) \text{cm}^{3}~\text{s}^{-1} \nonumber
\end{eqnarray}

ATMOS also encodes it as a two-body reaction, with rate law \citep{Sander2003}:
\begin{eqnarray}
    CO + OH \rightarrow CO_2 + H\nonumber\\
    k=1.5\times10^{-13} \text{cm}^{3}~\text{s}^{-1} \times(1.0 + 0.6\times P_{atm}) \nonumber
\end{eqnarray}
where $P_{atm}$ is the pressure in atmospheres. 

Kasting encodes it following \citet{Sander2011}. Note that the functional form linking $k_{inf}$ and $k_0$ to the rate constant $k$ is different than the standard expression for three-body reaction rates:

\begin{eqnarray}
    CO + OH \rightarrow CO_2 + H\nonumber\\
    k_0=1.5\times10^{-13}~\text{cm}^{3}~\text{s}^{-1}~(\frac{T}{300~K})^{0.6}\nonumber\\
    k_{\infty}=2.1\times10^{9}~\text{s}^{-1}~(\frac{T}{300~K})^{6.1}\nonumber\\
    k=(\frac{k_0}{1+\frac{k_0}{k_\infty/[M]}})0.6^{[1+(\log_{10}[\frac{k_0}{k_\infty/[M]}])^{2}]^{-1}} \nonumber\\
\end{eqnarray}

\begin{figure}[h]
\plotone{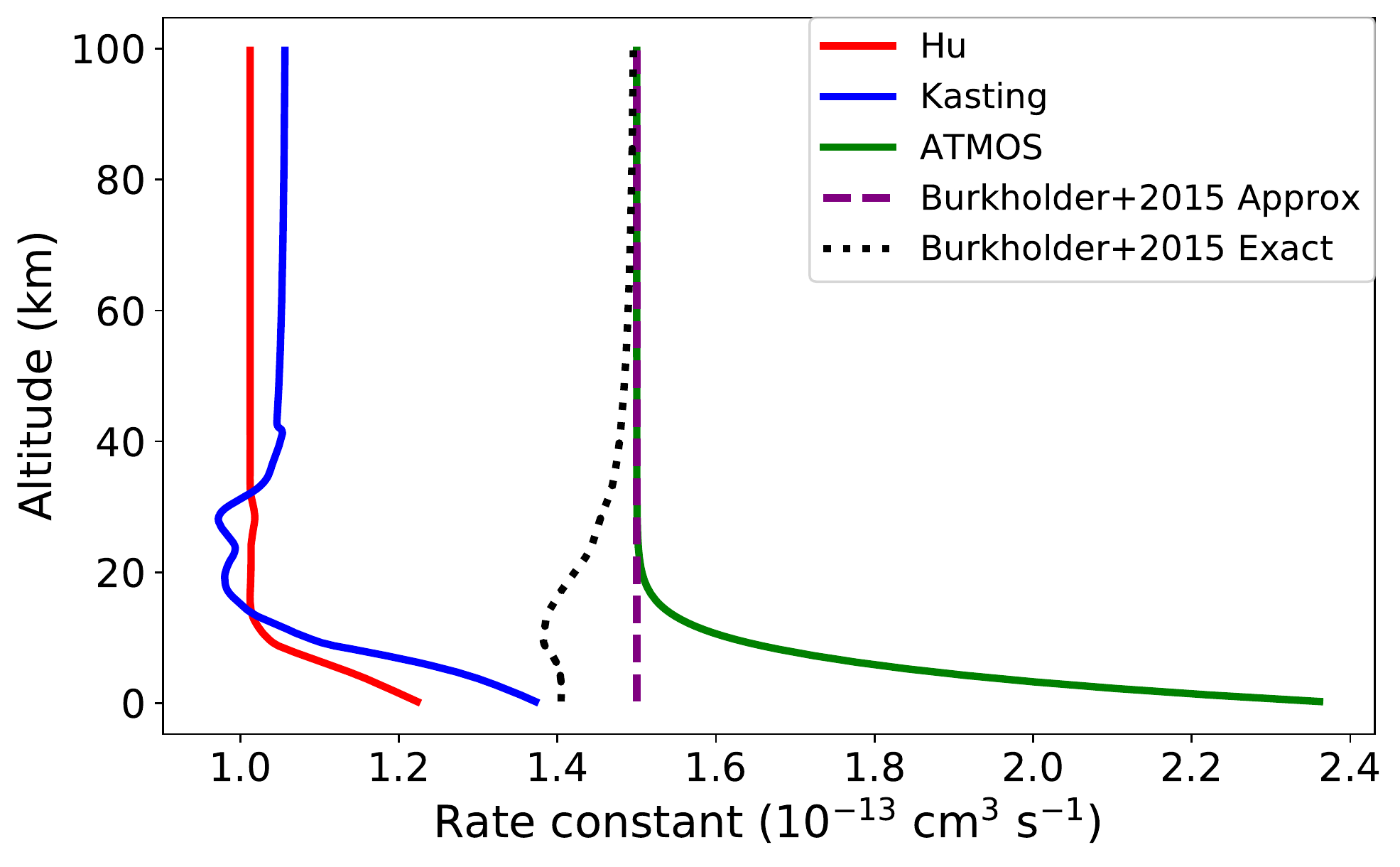}
\caption{Different effective bimolecular rate laws for $CO+OH\rightarrow CO_2 + H$ as a function of altitude in the T-P profile corresponding to this planetary scenario, calculated by ATMOS (see Figure~\ref{fig:tph2oeddy}). The lower atmosphere, where the H$_2$O photolysis rate is high and this reaction most important, is also where these rate laws most strongly disagree \label{fig:co-oh-ratelaws}}
\end{figure}

This differs slightly from the most recent JPL Chemical Kinetics Evaluation \citep{Burkholder2015}, where the exponent for the temperature dependence of $k_{0}$ is 0 rather than 0.6. For most terrestrial atmospheric applications, \citet{Burkholder2015} state that this reaction rate can be approximated as bimolecular, with rate constant $k\approx k_0=1.5\times10^{-13}~ \text{cm}^{3}~\text{s}^{-1}$. This formalism falls intermediate to the ATMOS and Hu/Kasting formalisms, agreeing with ATMOS in the upper atmosphere and with Hu/Kasting in the lower atmosphere. Note that since most H$_2$O and hence OH production is in the lower atmosphere, the lower atmosphere is photochemically overweighted. Figure~\ref{fig:co-oh-ratelaws} presents the differing CO+OH ratelaws used in our models as a function of altitude for the temperature-pressure profile considered in this study. In sum, variation in the $CO+OH$ ratelaw can affect $r_{CO}$ by $1.5-2\times$, and the most recently proposed ratelaw (from \citealt{Burkholder2015}) is intermediate to the ratelaws used to date. 

\subsubsection{S + CO Ratelaw}
ATMOS includes the reaction $S + CO + M\rightarrow OCS + M$, with rate law equal to that of $O + CO + M \rightarrow CO_2 + M$, following \citet{Mills1998}. This reaction has not been observed in the laboratory, but has been included as it is inferred from the atmospheric photochemistry of Venus, particularly the presence of OCS in its lower atmosphere \citep{Krasnopolsky2007, Yung2009}. Inclusion of this reaction decreases $r_{CO}$ by $20\%$, and supports $\sim$ppb levels of OCS. When this reaction is excluded, OCS essentially does not exist in the atmospheric scenarios we consider here. We identify this reaction as a key target for experimental characterization. 

\subsection{Atmospheric Profile\label{sec:atm_prof}}
\begin{figure}[h]
\epsscale{0.8}
\plotone{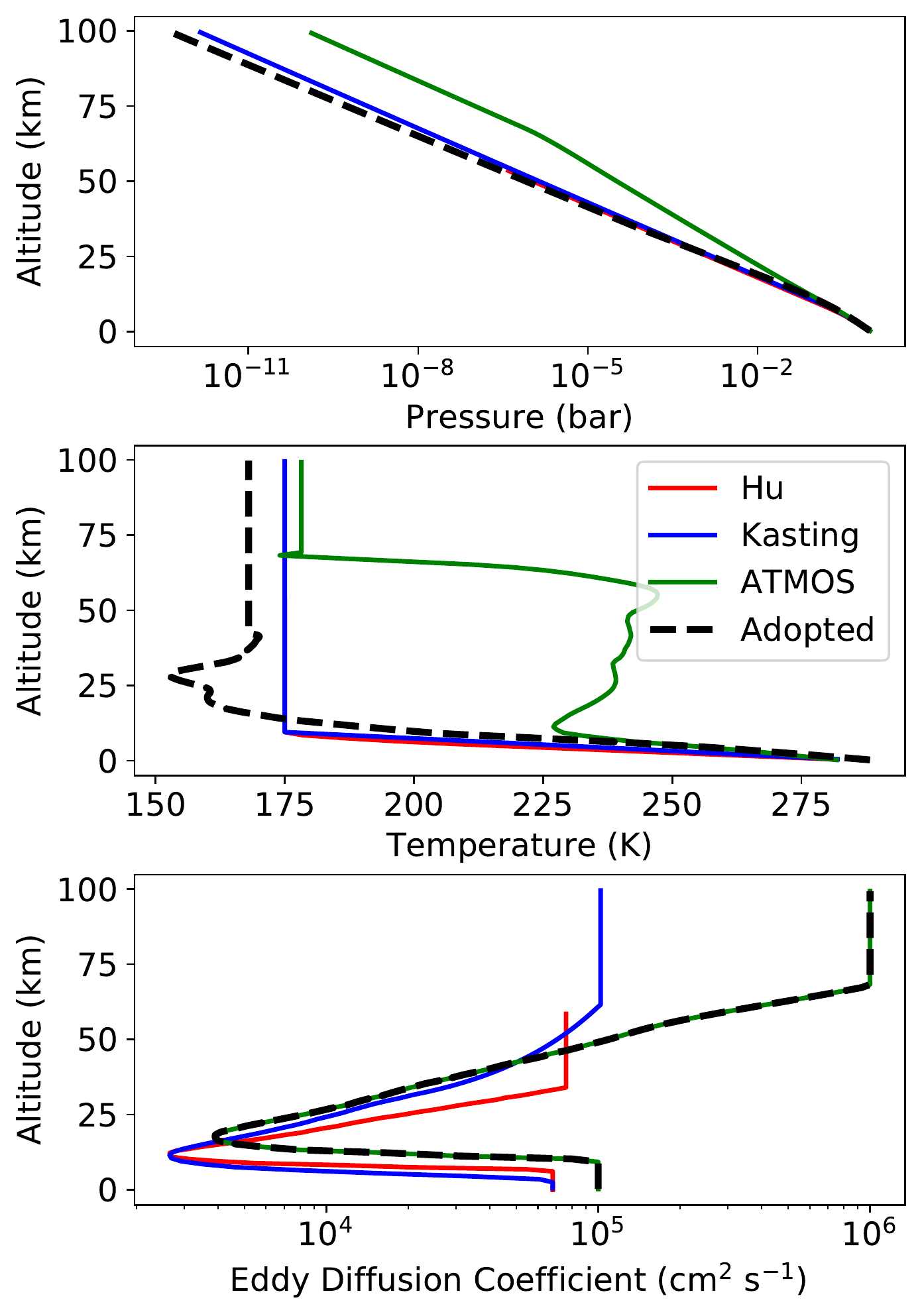}
\caption{Pressure, temperature, and eddy diffusion coefficients as a function of altitude in the baseline models at project outset (solid colored lines; \citealt{Hu2012, Harman2015, Arney2016}). The original ATMOS T-P profile was derived from an unconverged haze run. Dashed black lines give these profiles as self-consistently calculated by ATMOS for the particular planetary scenario we consider here; these are the profiles adopted for purposes of intercomparison (c.f. Section~\ref{sec:dance_models_dance}). It is crucial to include the cold stratosphere that results from the presence of CO$_2$ and the lack of shortwave absorbers; otherwise, the upper atmosphere is moistened and the H$_2$O photolysis rate is unphysically enhanced. \label{fig:tph2oeddy}}
\end{figure}

In anoxic, abiotic CO$_2$-rich atmospheres, the stratosphere is cold due to efficient line cooling by CO$_2$ and the absence of shortwave-absorbing stratospheric O$_3$ (from biogenic O$_2$) or haze (from biogenic CH$_4$) \citep{Kasting1984, Roble1995, Wordsworth2013CO2, Rugheimer2018}. Cold stratospheres are dry, due to the low saturation pressure of H$_2$O at low temperatures \citep{Wordsworth2013CO2}. If this fact is neglected (i.e. if the stratosphere is allowed to be relatively warm and moist), then vigorous H$_2$O photolysis in the wet upper atmosphere generates abundant OH which suppresses $r_{CO}$ regardless of assumptions on H$_2$O and CO$_2$ cross-sections. The baseline T/P profile in the ATMOS \texttt{Archaean+haze} template is warm ($\sim230$ K) because it was calculated for conditions in which shortwave absorption due to haze heats the stratosphere. By contrast, at the low CH$_4$/CO$_2$ ratios expected for CO$_2$-dominated abiotic exoplanets, haze formation is not expected, and the stratosphere should be cold ($\sim 150-170$ K) and dry \citep{DeWitt2009, Guzman-Marmolejo2013, Arney2016}. This means that when applying the \texttt{Archaean+haze} template from ATMOS to this planetary scenario, it is important to first re-calculate temperature-pressure profiles that are consistent with this scenario, e.g. by using the CLIMA module of ATMOS (Figure~\ref{fig:tph2oeddy}). Neglect of self-consistent climate can lead to overestimating upper-atmospheric [H$_2$O] (and hence H$_2$O photolysis rates) by 2-4 orders of magnitude.

\subsection{H and H$_2$ Escape Rates}
Atmospheric H$_2$ stabilizes CO by suppressing OH and O \citep{Kasting1983, Kharecha2005}. In abiotic, anoxic atmospheres, pH$_2$ is generally set by a balance between H$_2$ outgassing and H and H$_2$ escape from the atmosphere, with the escape velocities calculated by:
\begin{eqnarray}
    v_{diff} = D(\frac{1}{H_0} - \frac{1}{H}) \nonumber\\ 
\end{eqnarray}
where $D(X, Y)$ is the molecular diffusion coefficient of $X$ through $Y$ in cm$^2$s$^{-1}$, $H_0$ is the scale height for the bulk atmosphere at the escape altitude, $H$ is the scale height for the escaping component at the escape altitude, and $v_{diff}$ is the diffusion-limited escape velocity. 

\begin{figure}[h]
\plotone{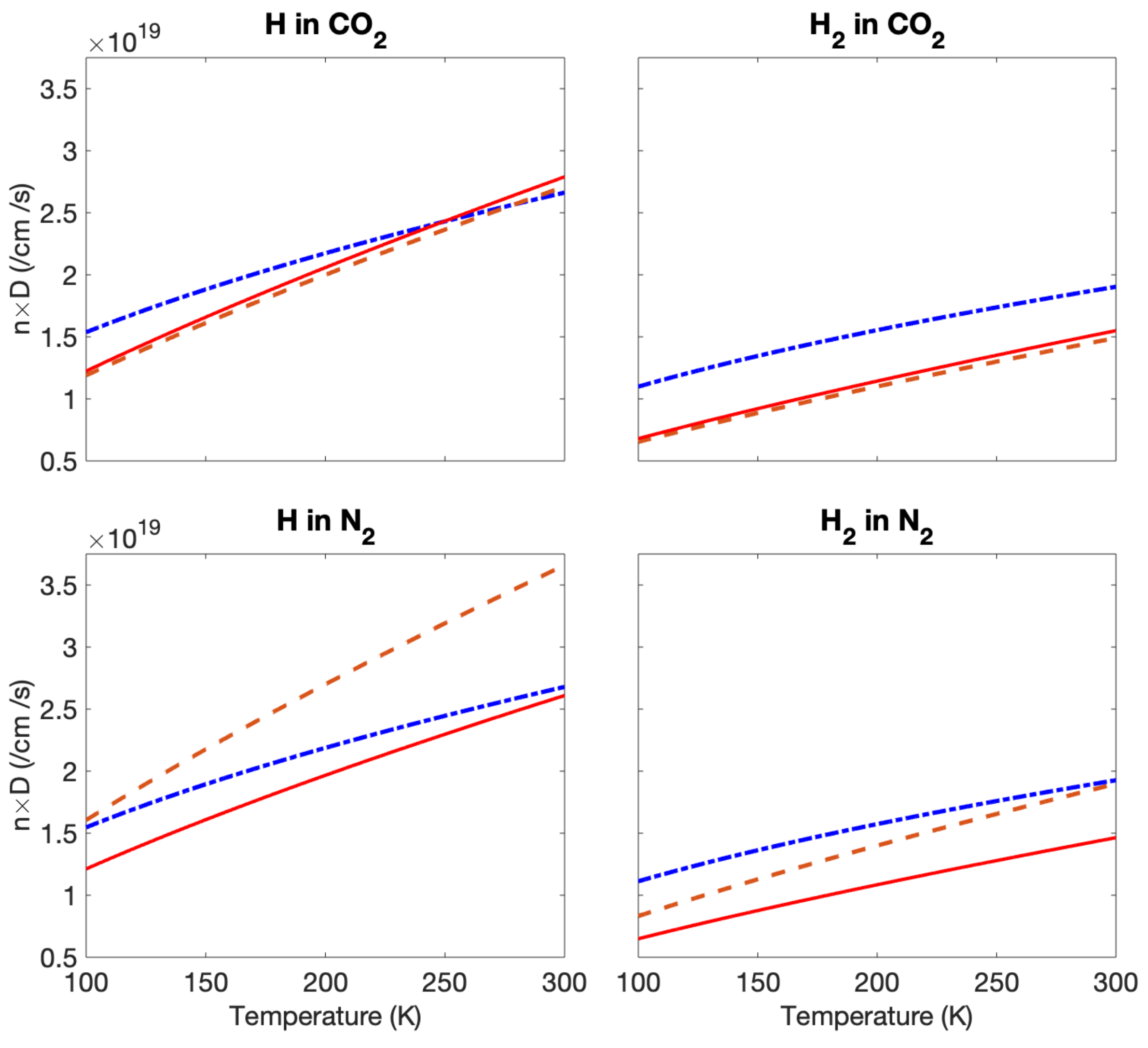}
\caption{$nD(X,Y)$ from different models. Red curves correspond to the Hu model, blue to the Kasting model, and orange to ATMOS. \label{fig:diffusion_differences}}
\end{figure}

The calculation of $D(X, Y)$ for H and H$_2$ through CO$_2$ and through N$_2$ are different between the Hu, Kasting, and ATMOS models (Figure~\ref{fig:diffusion_differences}). Specifically, the Kasting model uses a generalized diffusion coefficient formulation, valid for an interaction radius of $3\times10^{-8}$ cm \citep{Banks1973}: 
\begin{eqnarray}
    D_{i} = \frac{1.52\times 10^{18} \times (\frac{1}{\mu} + \frac{1}{\bar{\mu}})^{0.5}\times T^{0.5} \text{cm}^{-1}~\text{s}^{-1}}{n} \label{eqn:gen_dif}
\end{eqnarray}
where $\mu$ is the molecular weight of the individual species, $\bar{\mu}$ is the mean molecular weight of the atmosphere, $n$ is the number density in cm$^{-3}$, and $T$ is the temperature in K. 

The Hu model uses individualized expressions for $D(X,Y)$. $D(H, N_2)$ and $D(H_2,N_2)$ are taken from \citet{Banks1973}. $D(H_2, CO2)$ is taken from \citet{Marrero1972}, with the caveat that the exponential correction factor $\exp(\frac{-11.7 K}{T})$ is evaluated at 175K, corresponding to the assumed stratospheric temperature in the CO$_2$-dominated case from \cite{Hu2012}; this simplification leads to a $<4\%$ deviation from $150-300$ K. $D(H,CO_2)$ is taken as $1.8\times D(H_2, CO_2)$ following the observation of \citet{Zahnle2008} that $D(H, He)=1.8\times D(H_2, He)$: 

\begin{eqnarray}
    D(H, CO_2)=\frac{3.87\times10^{17}T^{0.75} \text{cm}^{-1}~\text{s}^{-1}}{n}\nonumber \\
    D(H_2, CO_2) = \frac{2.15\times10^{17}T^{0.75} \text{cm}^{-1}~\text{s}^{-1}}{n}\nonumber \\
    D(H, N_2)= \frac{4.87\times10^{17}T^{0.698} \text{cm}^{-1}~\text{s}^{-1}}{n} \nonumber \\
    D(H_2, N_2) = \frac{2.15\times10^{17}T^{0.740} \text{cm}^{-1}~\text{s}^{-1}}{n} \nonumber
\end{eqnarray}

The ATMOS model uses:
\begin{eqnarray}
    D(H, CO_2)=\frac{2.0\times10^{19}(T/200K)^{0.75}}{n}\\
    D(H_2, CO_2)=\frac{1.1\times10^{19}(T/200K)^{0.75}}{n}\\
    D(H, N_2)=\frac{2.7\times10^{19}(T/200K)^{0.75}}{n}\\
    D(H_2, N_2)=\frac{1.4\times10^{19}(T/200K)^{0.75}}{n}\\
\end{eqnarray}

Figure~\ref{fig:diffusion_differences} shows the variation in diffusion coefficient as a function of temperature for the formalisms selected by the different models. $nD$ does not vary significantly as a function of background gas for the Hu and Kasting models; however, $nD$ varies significantly as a function of background gas in ATMOS. By default, ATMOS uses coefficients for diffusion through N$_2$. Correcting these to the coefficients for diffusion through CO$_2$ (e.g., ATMOS ``Mars" setting) results in a $1.5\times$ increase in pH$_2$ and a $2\times$ increase for $r_{CO}$. For diffusion through CO$_2$, the Kasting formalism leads to higher escape of H$_2$ (and hence lower pH$_2$ and $r_{CO}$) compared to the Hu and ATMOS models. 

\subsection{Demonstrating Model Agreement\label{sec:dance_models_dance}}
To demonstrate that we have identified the key parameters driving the published variation in $r_{CO}$ between models, we run all three models under identical assumptions and show that our models agree and that we can reproduce both the high and low $r_{CO}$ published in the literature. Table~\ref{tab:high_and_low_co} enumerates the conditions required to achieve high and low CO in our models. Table~\ref{tab:constant_parameters} lists the other parameters that must be aligned between the models. We note that the choices of the constant parameters (Table~\ref{tab:constant_parameters}) are chosen primarily to facilitate the numerical experiment of the intercomparison, not for physical realism. For example, we essentially neglect rainout for purposes of this intercomparison; we do this not because CO$_2$-dominated rocky planets should lack rain, but because this is the most tractable rainout regime we can force all three of our models into, and because the results are not sensitive to this assumption. Similarly, when considering CO$_2$ absorption, we use cross-sections from the Kasting/ATMOS models, not because they are the most current estimates of CO$_2$ absorption, but because it is easier to incorporate these cross-sections into the Hu model than vice versa. Finally, we note that while that mixing length theory suggests the CO$_2$-dominated atmospheres should be more turgid than N$_2$-dominated atmospheres \citep{Hu2012, Harman2015}, using the \citet{Hu2012} eddy diffusion profile for CO$_2$-dominated atmospheres leads to numerical instabilities in the ATMOS and Kasting models. We consequently elect to use the ATMOS eddy diffusion profile, calibrated for N$_2$-dominated atmospheres, in this numerical experiment. It is unclear why reduced eddy diffusion should lead to such instabilities; we intend further studies to answer this question. Overall, we therefore emphasize that the use of these parameters should not necessarily constitute an endorsement, as they were primarily chosen to facilitate model intercomparison. The models were corrected for the errors summarized in Table~\ref{tab:corrections}, and the simulation parameters were otherwise as given in Tables~\ref{tab:detailed_parameters} and ~\ref{tab:species_bcs}.

\begin{deluxetable}{ccc}[h]
\tablecaption{Parameters defining high and low CO cases (\citet{Hu2012} CO$_2$-dominated scenario) file\label{tab:high_and_low_co}}
\tablewidth{0pt}
\tablehead{\colhead{Scenario Parameter} &\colhead{High CO} & \colhead{Low CO}}
\startdata
H$_2$O cross-sections (ATMOS/Kasting)& Terminated at 198 nm  & Extrapolated to 208.3 nm \\
H and H$_2$ Diffusion Coefficient & Hu/ATMOS (CO$_2$-dominated) & Kasting \\
CO + OH Ratelaw & Hu/Kasting & ATMOS \\
S + CO & No & Yes$^*$ 
\enddata
\tablecomments{$^*$The Kasting model \citep{Harman2015} does not include OCS, and hence does not include this reaction}
\end{deluxetable}

\begin{deluxetable}{cccc}[h]
\tablecaption{Parameters held constant between high and low CO cases\label{tab:constant_parameters}}
\tablewidth{0pt}
\tablehead{\colhead{Scenario Parameter} & \colhead{Hu} & \colhead{ATMOS} & \colhead{Kasting}}
\startdata
Semimajor Axis & & 1.21 AU & \\
TP Profile &  & Calculated from ATMOS (Fig.~\ref{fig:tph2oeddy}) & \\
Eddy Diffusion &  & ATMOS (Fig.~\ref{fig:tph2oeddy}) & \\
Surface Albedo &  & 0.25 & \\
CO$_2$ Cross-Sections &  & As in Kasting/ATMOS \citep{Kasting1981} & \\ 
Vertical Resolution &  & $0-100$ km, 1 km resolution & \\
Lightning &  & Off & \\
Rainout &  & $10^{-9}\times$ Earthlike & \\
Global Redox Balance Enforced & No & No & No/Yes\\
\enddata
\end{deluxetable}

The results of this numerical experiment are given in Table~\ref{tab:intercomparison_results}. Our models (1) agree with each other to within a factor of 2, and (2) can reproduce both the low ($\sim$200 ppm) and high ($\sim8200$ ppm) CO surface mixing ratios published in the literature (e.g., \citealt{Hu2012, Harman2015}). Agreement between ATMOS and Hu is within $10\%$, reflecting particularly intensive intercomparison. We conclude that we have successfully identified the parameters driving divergent predictions of $r_{CO}$ in this planetary scenario.

\begin{deluxetable}{cccc}[h]
\tablecaption{CO surface mixing ratios for the three models, applied to our abiotic CO$_2$-N$_2$ atmospheric scenario. The first line presents the mixing ratios from the models at the outset of the project, i.e. prior to fixing the errors in Table~\ref{tab:corrections}. $r_{CO}$ disagreed by $50\times$. The second and third line present the mixing ratios calculated from the models after fixing the errors, with uniform simulation parameters, under assumptions corresponding to both high and low CO (Tables~\ref{tab:constant_parameters}, \ref{tab:high_and_low_co}). Calculated $r_{CO}$ agreed within a factor of $2\times$ between models, demonstrating that we had successfully identified and accounted for the factors driving divergent $r_{CO}$ between models\label{tab:intercomparison_results}}
\tablewidth{0pt}
\tablehead{\colhead{Scenario Parameter} &\colhead{$r_{CO}$ (Hu)}  & \colhead{$r_{CO}$ (ATMOS)}   &  \colhead{$r_{CO}$ (Kasting)} }
\startdata
Initially & 8.2E-3 & 1.72E-4  & 2.1E-4 \\
\hline
Low-CO Assumptions  & 2.7E-4 &  2.6E-4 &   1.3E-4/1.5E-4$^{*}$ \\
High-CO Assumptions & 5.9E-3 & 5.4E-3 & 8.1E-3/8.2E-3$^{*}$ \\
\enddata
\tablecomments{$^{*}$Global redox balance enforced as per \citet{Harman2015}.}
\end{deluxetable}

\clearpage
\bibliography{photochem-co}{}
\bibliographystyle{aasjournal}
https://www.overleaf.com/project/5dd5a1ef1026ed0001d29c1b


\end{document}